\documentclass[%
 reprint,
 superscriptaddress,
 amsmath,
 amssymb,
 aps,
]{revtex4-1}

\usepackage{graphicx}% Include figure files
\usepackage{dcolumn}% Align table columns on decimal point
\usepackage{bm}% bold math

\usepackage{hyperref}% add hypertext capabilities

\newcommand{\argmax}{\mathop{\rm arg~max}\limits}

\newcommand{\e}{\mathrm{e}}

\begin{document}

\title{Extracting phase coupling functions between collectively oscillating networks directly from time-series data}
% \title{Extracting phase coupling functions between networks of dynamical elements that exhibit collective oscillations: Direct extraction from time-series data}

\author{Takahiro Arai}
\email{arai@acs.i.kyoto-u.ac.jp}
\affiliation{Graduate School of Informatics, Kyoto University, Yoshida-Honmachi, Sakyo-ku, Kyoto 606-8501, Japan}

\author{Yoji Kawamura}
\affiliation{Center for Mathematical Science and Advanced Technology, Japan Agency for Marine-Earth Science and Technology, Yokohama 236-0001, Japan}

\author{Toshio Aoyagi}
\affiliation{Graduate School of Informatics, Kyoto University, Yoshida-Honmachi, Sakyo-ku, Kyoto 606-8501, Japan}

\date{\today}% It is always \today, today,

\begin{abstract}
Many real-world systems are often regarded as weakly coupled limit-cycle oscillators,
in which each oscillator corresponds to a dynamical system with many degrees of freedom that have collective oscillations.
One of the most practical methods for investigating the synchronization properties of such a rhythmic system is
to statistically extract phase coupling functions between limit-cycle oscillators directly from observed time-series data.
In Particular, using a method that combines phase reduction theory and Bayesian inference,
the phase coupling functions can be extracted from the time-series data
of even just one variable in each oscillatory dynamical system with many degrees of freedom.
However,
it remains unclear
how the choice of the observed variables affects the statistical inference for the phase coupling functions.
In this study,
we examine the influence of observed variable types on the extraction of phase coupling functions
using some typical dynamical elements under various conditions.
We demonstrate that
our method can consistently extract the macroscopic phase coupling functions
between two phases representing collective oscillations in a fully locked state, regardless of the observed variable types;
for example, even using one variable of any element in one system and the mean-field value over all the elements in another system.
We also study the case of globally coupled phase oscillators in a partially locked state.
Our results reveal directional asymmetry in the robustness of extracting the macroscopic phase coupling function between two networks.
For instance, 
when an asynchronous oscillator in network $A$ and the macroscopic collective oscillation of network $B$ is observed,
the macroscopic phase coupling function from network $A$ to network $B$ can be extracted more robustly than in the opposite direction.
\end{abstract}

\maketitle

%\tableofcontents

%% intro %%%%%%%%%%%%%%%%%%%%%%%%%%%%%%%
\section{Introduction}
\label{sec:intro}

\begin{figure*}
  \includegraphics[width=0.9\textwidth]{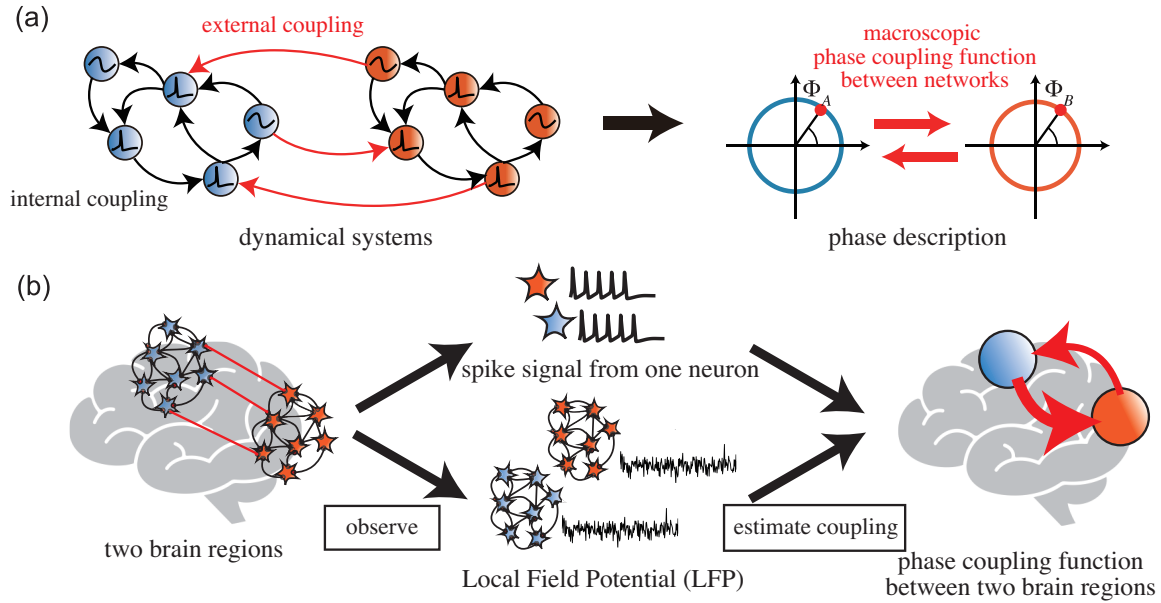}
  \caption{
    Schematic diagrams of
    the phase reduction method for collective oscillation
    and statistical inference of the phase coupling function.
    {\bf (a)}~
    Schematic diagram of the phase reduction method for collective oscillation.
    The left part of the figure shows that
    each network has many elements,
    and internal couplings (black arrows) and external couplings (red arrows) exist.
    When each network has a stable limit-cycle and the external couplings are sufficiently weak,
    the dynamics of each network can be projected onto a single phase equation, as illustrated in the right part of the figure.
    The macroscopic phase coupling function between the networks (bold red arrows) can be derived as a simple form.
    {\bf (b)}~
    Schematic diagram of statistical inference of the connectivity between brain regions.
    The dynamics of one brain region,
    which is a neural population,
    can be observed as
    spike signals of one neuron,
    local field potential (LFP),
    and so on.
    The phase coupling function is extracted from observed time-series,
    however,
    the same results may not necessarily be obtained from two different observation methods.
  }
  \label{fig:intro}
\end{figure*}

\par
Synchronization phenomena of coupled dynamical elements are known in many fields,
such as biological and engineering systems,
and often play important roles~\cite{kuramoto_chemical_1984}.
Further,
collective synchronization has been widely studied
for both globally coupled systems and complex network systems~\cite{
  strogatz_kuramoto_2000, acebron_kuramoto_2005, arenas_synchronization_2008, dorogovtsev_critical_2008}.
The framework to derive a low-dimensional description of the dynamics has been developed to analyze the collective dynamics.
One of the most successful and widely used theoretical methods is phase reduction theory~\cite{
  winfree_biological_1967, winfree_geometry_2001,
  ermentrout_multiple_1991, ermentrout_stable_1992, ermentrout_type_1996,
  pikovsky_finite-size_1999, pikovsky_synchronization_2001,
  brown_phase_2004, izhikevich_dynamical_2007, ashwin_mathematical_2016, nakao_phase_2016}.
This method derives the phase description,
in which the dynamics in the vicinity of the limit-cycle is projected onto a single phase equation,
such that the interaction between oscillators can be simply described as a phase coupling function.

\par
Recently,
statistical methods to extract the phase coupling function directly from observed time-series have been developed:
methods to reconstruct
the phase dynamics of two oscillators~\cite{
  kralemann_uncovering_2007, kralemann_phase_2008, kralemann_vivo_2013},
effective connectivity of an oscillator network~\cite{
  kralemann_reconstructing_2011, kralemann_reconstructing_2014},
phase coupling function~\cite{
  kiss_predicting_2005, tokuda_inferring_2007, miyazaki_determination_2006}, and
phase sensitivity function~\cite{
  galan_efficient_2005, ermentrout_relating_2007, ota_weighted_2009, imai_improvement_2016, funato_evaluation_2016}.
Further,
several methods using the Bayesian inference framework~\cite{bishop_pattern_2006} have been proposed
to extract the phase description~\cite{
  stankovski_inference_2012, duggento_dynamical_2012, stankovski_coupling_2017, stankovski_time-varying_2017,
  hagos_synchronization_2019, ota_direct_2014}.
These methods have been used to analyze various systems, such as the
choruses of frogs~\cite{
  ota_interaction_2020},
electroencephalography data~\cite{
  stankovski_coupling_2015, stankovski_alterations_2016, stankovski_neural_2017, onojima_dynamical_2018}, and
spiking neurons~\cite{
  suzuki_bayesian_2018}.

\par
In general,
real-world systems are comprised of many interacting subsystems,
in which each subsystem is also comprised of many elements and exhibits collective dynamics.
These subsystems often interact with each other to coordinate their functional activities.
For example,
each brain region exhibits synchronous oscillation dynamics of the neural population,
and the synchronization of these oscillations between brain regions
has been studied as the effective connectivity~\cite{
  sakkalis_review_2011, jafarian_structure_2019, kralemann_reconstructing_2011, kralemann_reconstructing_2014, stankovski_coupling_2021}.
The effective connectivity has been widely applied to study the
resting-state of the human brain~\cite{ponce-alvarez_resting-state_2015},
epileptic seizures~\cite{rosch_calcium_2018, van_mierlo_network_2019}, and
cognition and working memory~\cite{park_structural_2013}.
Both theory and inference methods for coupling functions between subsystems are significant in neuroscience,
because the brain's architecture is a highly connected complex system~\cite{stankovski_coupling_2021}.

\par
Here,
we consider a dynamical system that consists of networks of dynamical elements.
The model reduction for collective dynamics has been intensively investigated
for analyzing the macroscopic synchronization properties between networks that show collective dynamics~\cite{
  pikovsky_dynamics_2015,
  pikovsky_partially_2008, pikovsky_dynamics_2011,
  watanabe_integrability_1993, watanabe_constants_1994,
  ott_low_2008, ott_long_2009}.
Moreover,
several theoretical frameworks to reduce the collective dynamics to a single phase variable have been developed~\cite{
  kawamura_collective_2008, kawamura_collective_2014, kawamura_collective_2017,
  kawamura_phase_2010-1, kawamura_phase_2010,
  kori_collective-phase_2009, kawamura_phase_2014, nakao_phase_2018}.
Therefore, using these methods, 
we can investigate a macroscopic phase sensitivity function
and macroscopic phase coupling function between networks.
Figure~\ref{fig:intro}(a) illustrates the framework developed in Ref.~\cite{nakao_phase_2018},
where two networks with collective oscillations have many internal and external couplings.
In this case,
the dynamics of the entire network is reduced to a one-dimensional phase equation,
and the derived macroscopic phase coupling function describes 
the effect of all external couplings between two networks on the phase dynamics.
Although analytical studies have been conducted as mentioned before,
inference methods for the collective dynamics have not yet been fully investigated.

\par
In this study,
to extract the macroscopic phase coupling functions between networks,
we extend the range of applications of the Bayesian inference method~\cite{ota_direct_2014}
from a phase representing the state of one element
to a macroscopic phase representing the state of an entire network
with collective oscillations~\cite{nakao_phase_2018, ott_low_2008, ott_long_2009, kawamura_phase_2010}.
The phase coupling function can be extracted from the time-series of only one variable
by introducing the phase description into the statistical inference method.
Note that it is sufficient to observe the time-series of only one variable of each network with collective oscillation, 
not all the variables.
However, although many methods to reconstruct the phase dynamics have been proposed,
it has not yet been clear whether the same phase dynamics can be obtained from the time-series of any component of a dynamical system.
In the case of real-world high-dimensional dynamical systems with collective oscillations,
it is unclear whether the selection of the component affects the reconstructed phase dynamics.

\par
Let us consider an example shown in Fig.~\ref{fig:intro}(b),
where the purpose is to analyze the macroscopic phase coupling functions between two brain regions from observed time-series data.
Most studies have supposed that the macroscopic phase coupling functions are extracted from the local field potential (LFP) time-series,
which reflects the activities of neural populations.
However,
it is unclear what the coupling function extracted from the time-series of the spike signals of one neuron represents, 
i.e., it is highly nontrivial to determine whether the coupling function estimated from the time-series of the spike signals of one neuron is the same as that estimated from the LFP time-series.
Moreover,
the extraction of coupling functions may also be affected by the choice of the elements for observation,
e.g., an excitatory or inhibitory neuron can be chosen.
From these viewpoints,
we evaluate the effect of the choice of observed variables on the extraction of the macroscopic phase coupling function
for two types of collective oscillations:
fully locked state~\cite{nakao_phase_2018} and partially locked state~\cite{kawamura_phase_2010}.
The fully locked state is the simplest case
to extract the macroscopic phase coupling function.
In this case, all the elements in a network follow the macroscopic oscillation.
However, it is uncertain whether the same phase coupling function could be extracted from any time series of observations,
because each element of the network has a different response to perturbations.
The partially locked state is a more complex case
to extract the macroscopic phase coupling function,
because there exist asynchronous oscillators.

\par
This paper is organized as follows.
In Sec.~\ref{sec:method},
we briefly review the phase reduction method for collective oscillation in the fully locked state 
and the Bayesian inference method for extracting the phase description directly from time-series data.
In Sec.~\ref{subsec:case1} to Sec.~\ref{subsec:case3},
we focus on networks where all the elements have collective oscillation in the fully locked state.
We apply the method to extract the interaction between networks for
two networks of FitzHugh-Nagumo (FHN) elements (Sec.~\ref{subsec:case1}),
three networks of FHN elements (Sec.~\ref{subsec:case2}), and
two networks of van der Pol oscillators (Sec.~\ref{subsec:case3}).
In these cases,
three types of observed variables are studied:
one excitable element, one oscillatory element, and a mean-field.
Further, we also investigate a case of two networks of globally coupled phase oscillators in a partially locked state (Sec.~\ref{subsec:case4}).
In this case,
we examine the possibility of extracting the macroscopic phase coupling function
using time-series data of an asynchronous oscillator.
In Sec.~\ref{sec:discussion},
we discuss the implications of the method for characterizing the interaction between networks directly from time-series data.

%% method %%%%%%%%%%%%%%%%%%%%%%%%%%%%%%%
\section{Methods}
\label{sec:method}

In this section,
we briefly review both the phase description for collective oscillations in the fully locked state (Sec.~\ref{subsec:phasedescription})
and the Bayesian framework to extract the collective phase description directly from time-series data (Sec.~\ref{subsec:Bayes}).

\subsection{Phase description}
\label{subsec:phasedescription}

We consider $N$ networks of coupled dynamical elements and weak interactions between the networks.
The dynamics of element $i~(i = 1, 2, \ldots, N_{\gamma})$ in network $\gamma$ is given by
\begin{eqnarray}
  \frac{d}{dt} \bm{X}_i^{\gamma}(t) = 
  && \bm{F}_i^{\gamma}( \bm{X}_i^{\gamma}(t) )
  + \sum_{j \neq i}^{N_{\gamma}}
  \bm{G}_{i j}^{\gamma}( \bm{X}_i^{\gamma}(t), \bm{X}_j^{\gamma}(t) )
  \nonumber \\
  && + \epsilon \sum_{\nu \neq \gamma} \sum_{j = 1}^{N_{\nu}}
  \bm{H}_{i j}^{\gamma \nu}( \bm{X}_i^{\gamma}(t), \bm{X}_j^{\nu}(t) )
  + \bm{\eta}_i^{\gamma}(t),
  \label{equ:general_form}
\end{eqnarray}
where
$\bm{X}_i^{\gamma}(t) \in {\bf R}^{d_i^{\gamma}}$
represents the $d_i^{\gamma}$-dimensional state of element $i$ in network $\gamma$ at time $t$,
$\bm{F}_i^{\gamma}: {\bf R}^{d_i^{\gamma}} \rightarrow {\bf R}^{d_i^{\gamma}}$
represents the individual dynamics of element $i$ in network $\gamma$, and
$\bm{G}_{i j}^{\gamma}: {\bf R}^{d_i^{\gamma}} \times {\bf R}^{d_j^{\gamma}} \rightarrow {\bf R}^{d_i^{\gamma}}$
represents the internal coupling from element $j$ to element $i$ in network $\gamma$.
We assume that self-coupling does not exist or is absorbed into the individual dynamics term $\bm{F}_i^{\gamma}$,
i.e., $\bm{G}_{i i}^{\gamma} = \bm{0}$.
In this system,
there are external couplings between elements belonging to different networks as
$\bm{H}_{i j}^{\gamma \nu}: {\bf R}^{d_i^{\gamma}} \times {\bf R}^{d_j^{\nu}} \rightarrow {\bf R}^{d_i^{\gamma}}$.
The intensity of the external couplings is determined by $\epsilon$.
Independent white Gaussian noise $\bm{\eta}_i^{\gamma} \in {\bf R}^{d_i^{\gamma}}$
is given to each element in each network.

\par
We introduce the phase description into the dynamical system described by Eq.~(\ref{equ:general_form}).
Here,
we assume that each network has collective oscillation in a fully locked state,
and that there are no perturbations to them,
i.e., $\epsilon = 0$ and $\bm{\eta}_i^{\gamma} = \bm{0}$.
Under this condition,
the dynamics of element $i$ in network $\gamma$ is assumed to exhibit periodic oscillation as follows:
\begin{eqnarray}
  \bm{X}_i^{\gamma}(t) = \bm{X}_i^{\gamma}(t + T_{\gamma}),
  \label{equ:collective_oscillation}
\end{eqnarray}
where each network possesses a stable limit-cycle solution.
In this situation,
all the elements in network $\gamma$ exhibit periodic oscillation with the period $T_{\gamma}$.
Let us consider the phase variable $\Phi_{\gamma}(t) \in [0, 2\pi)$.
The state of network $\gamma$ at time $t$ can be described as a function of the phase variable,
i.e., $\bm{X}_i^{\gamma}(t) = \bm{X}_i^{\gamma}( \Phi_{\gamma}(t) )$,
where the value of $\Phi_{\gamma}$ indicates the state of network $\gamma$,
which is in the vicinity of the limit-cycle.
Without any perturbations,
the phase variable, $\Phi_{\gamma}(t)$, is expected to increase with a constant natural frequency $\Omega_{\gamma}$ as follows:
\begin{eqnarray}
  \frac{d}{dt} \Phi_{\gamma}(t) = \Omega_{\gamma},
  \label{equ:natural_frequency}
\end{eqnarray}
where $\Omega_{\gamma} := 2\pi / T_{\gamma}$.

\par
We now consider a case in which perturbation to each network is given
but the perturbation intensity is so weak that the collective oscillation persists.
To satisfy this condition,
the intensity of external couplings between networks is set to $\epsilon \ll 1$,
and the noise intensity is also assumed to be small.
The effect of the perturbation appears on the phase variable in this situation.
The dynamics of each network described by Eq.~(\ref{equ:general_form})
can be reduced to the following phase equation using the phase description~\cite{nakao_phase_2018}:
\begin{eqnarray}
  \frac{d}{dt} \Phi_{\gamma}(t)
  = \Omega_{\gamma}
  + \epsilon \sum_{\nu \neq \gamma} \Gamma_{\gamma \nu}( \Delta \Phi_{\gamma \nu} )
  + \xi_{\gamma}(t),
  \label{equ:phase_description}
\end{eqnarray}
where $\Delta \Phi_{\gamma \nu} := \Phi_{\nu} - \Phi_{\gamma}$ is the phase difference between networks $\gamma$ and $\nu$, and
$\Gamma_{\gamma \nu}(\Delta \Phi_{\gamma \nu})$ is the phase coupling function from network $\nu$ to network $\gamma$.
The phase coupling function $\Gamma_{\gamma \nu}$ is obtained by averaging the product of phase sensitivity function
and external coupling $\bm{H}_{i j}^{\gamma \nu}$ over the period of the collective oscillation~\cite{nakao_phase_2018}.
The natural frequency of the collective oscillation, $\Omega_{\gamma}$, is the product of phase sensitivity function
and the summation of the first and second terms in Eq.~(\ref{equ:general_form}).
The independent white Gaussian noise $\xi_{\gamma}(t)$ given to network $\gamma$ satisfies
$\langle \xi_{\gamma}(t) \rangle = 0$ and
$\langle \xi_{\gamma}(t) \xi_{\nu}(s) \rangle = 2 D_{\gamma} \delta_{\gamma \nu} \delta(t - s)$,
where $\delta_{\gamma \nu}$ and $\delta(t)$ are the Kronecker delta and Dirac delta functions,
respectively.
The phase coupling function, $\Gamma_{\gamma \nu}(\Delta \Phi_{\gamma \nu})$,
which depends only on the phase difference, $\Delta \Phi_{\gamma \nu}$,
represents the effect of all the external couplings,
$\bm{H}_{i j}^{\gamma \nu}( \bm{X}_i^\gamma(\Phi_\gamma), \bm{X}_j^\nu(\Phi_\nu) )$,
from network $\nu$ to network $\gamma$.
The noise, $\xi_{\gamma}(t)$, represents the effect of all $\bm{\eta}_i^{\gamma}(t)$ given to network $\gamma$.

\par
Because the phase coupling function $\Gamma_{\gamma \nu}(\Delta \Phi_{\gamma \nu})$ is a $2\pi$-periodic function,
it can be expanded as a Fourier series
\begin{eqnarray}
  \Gamma_{\gamma \nu}(\Delta \Phi_{\gamma \nu})
  = a_{\gamma \nu}^{(0)}
  + \sum_{m = 1}^{M_{\gamma \nu}} \Big(
  && a_{\gamma \nu}^{(m)} \cos(\Delta \Phi_{\gamma \nu})
  \nonumber \\
  && + b_{\gamma \nu}^{(m)} \sin(\Delta \Phi_{\gamma \nu}) \Big),
  \label{equ:phase_coupling}
\end{eqnarray}
where the phase coupling function, $\Gamma_{\gamma \nu}(\Delta \Phi_{\gamma \nu})$,
has been expanded up to $M_{\gamma \nu}$th order.
The maximum order $M_{\gamma \nu}$ is determined from the model selection (Sec.~\ref{subsec:Bayes}).
To eliminate the redundancy of the two constant terms, $a_{\gamma \nu}^{(0)}$ and $\Omega_\gamma$,
we define $\hat{\Omega}_\gamma := \Omega_\gamma + \epsilon \sum_{\nu \ne \gamma} a_{\gamma \nu}^{(0)}$ and
$\hat{\Gamma}_{\gamma \nu}(\Delta \Phi_{\gamma \nu}) := \Gamma_{\gamma \nu}(\Delta \Phi_{\gamma \nu}) - a_{\gamma \nu}^{(0)}$.

\par
The Bayesian inference method can be applied to various cases by adopting appropriate basis functions.
The method for generalized phase equation form was earlier studied
in Refs.~\cite{stankovski_inference_2012, duggento_dynamical_2012}.
In this study, we adopt the Kuramoto--Daido form~\cite{kuramoto_chemical_1984, daido_onset_1996},
which can also be extended to various cases.
For instance, we can adopt the $n:m$ phase coupling function to evaluate $n:m$ synchronization~\cite{onojima_dynamical_2018}.
We will discuss the advantage of using the Kuramoto--Daido form as the phase coupling function in Sec.~\ref{sec:discussion}.

\subsection{Extraction of phase description directly from time-series data}
\label{subsec:Bayes}

The Bayesian inference method~\cite{ota_direct_2014} is used
to extract the phase dynamics described by Eq.~(\ref{equ:phase_description})
directly from phase time-series data $\{\Phi_{\gamma}(t_l)\}_{\gamma, l}$~$(l = 0, 1, \ldots, L)$.
In the model identification for the phase dynamics of network $\gamma$,
we estimate the parameters
$\hat{\Omega}_\gamma$,
$\{ a_{\gamma \nu}^{(1)}, b_{\gamma \nu}^{(1)}, \ldots, a_{\gamma \nu}^{(M_{\gamma \nu})}, b_{\gamma \nu}^{(M_{\gamma \nu})} \}_{\nu}$, and
$\hat{D}_{\gamma} := 2D_{\gamma} / \Delta t$,
where $\Delta t$ is the sampling interval of the phase time-series.
For simplicity, we denote the phase time-series as $\{\Phi (t)\}$ instead of $\{\Phi_{\gamma}(t_l)\}_{\gamma, l}$,
and we also use the following shorthand notation:
\begin{eqnarray}
  && \bm{c}_{\gamma} := 
  ( \hat{\Omega}_{\gamma}, \bm{c}_{\gamma, 1}, \ldots, \bm{c}_{\gamma, \gamma-1}, \bm{c}_{\gamma, \gamma+1}, \ldots, \bm{c}_{\gamma, N} )^{\mathrm{T}},
  \nonumber \\
  && \bm{c}_{\gamma, \nu} := 
  ( a^{(1)}_{\gamma \nu}, b^{(1)}_{\gamma \nu}, \ldots, a^{(M_{\gamma \nu})}_{\gamma \nu}, b^{(M_{\gamma \nu})}_{\gamma \nu} ).
  \nonumber
\end{eqnarray}
The notation $\bm{c}_{\gamma}$ refers to the coefficients of all basis functions
in Eqs.~(\ref{equ:phase_description}) and (\ref{equ:phase_coupling}),
while $\bm{c}_{\gamma, \nu}$ refers to the coefficients of the basis functions
in the coupling function $\hat{\Gamma}_{\gamma, \nu}$.

\par
We define the likelihood function as follows:
\begin{eqnarray}
  && P\left( \{\Phi(t) \} \middle| \bm{c}_{\gamma}, \hat{D}_{\gamma} \right)
  \nonumber \\
  && = \prod_{l=0}^{L-1} \mathcal{N}\left( \dot{\Phi}_{\gamma}(t_l) \middle|
  \hat{\Omega}_{\gamma} + \epsilon \sum_{\nu \neq \gamma} \hat{\Gamma}_{\gamma \nu}(\Delta \Phi_{\gamma \nu}(t_l)),
  \hat{D}_{\gamma} \right),
  \label{equ:likelihood}
\end{eqnarray}
where the Gaussian distribution corresponds to model fitting of the nonlinear equation,
i.e., Eq.~(\ref{equ:phase_description}).
The phase derivative is evaluated using the first-order forward difference, i.e.,
$\dot{\Phi}_{\gamma}(t_l) := \frac{\Phi_{\gamma}(t_{l+1}) - \Phi_{\gamma}(t_l)}{t_{l+1} - t_l}$.
We adopt a Gaussian-inverse-gamma distribution for the conjugate prior distribution, 
as follows:
\begin{eqnarray}
  P_{\mathrm{prior}}( \bm{c}_{\gamma}, \hat{D}_{\gamma} )
  = \mathcal{N}\left( \bm{c}_{\gamma} \middle| \bm{\chi}_0, \hat{D}_{\gamma} \Sigma_0 \right)
  \mathcal{IG}\left( \hat{D}_{\gamma} \middle| \alpha_0, \beta_0 \right),
  \label{equ:prior}
\end{eqnarray}
where $\bm{\chi}_0$, $\Sigma_0$, $\alpha_0$, and $\beta_0$ are the hyperparameters.
The mean and covariance matrix of the Gaussian distribution in Eq.~(\ref{equ:prior})
are determined by $\bm{\chi}_0$ and $\Sigma_0$, respectively,
while the shape parameter and scale parameter of the inverse-gamma distribution
are $\alpha_0$ and $\beta_0$, respectively.
In this study,
we consider $\bm{\chi}_0 := \bm{0}$ and $\Sigma_0 := \lambda_0^{-1} {\bf I}$,
where ${\bf I}$ is an identity matrix.
We can easily calculate the posterior distribution of $\bm{c}_{\gamma}$ and $\hat{D}_{\gamma}$ as follows:
\begin{eqnarray}
  P_{\mathrm{post}}\left( \bm{c}_{\gamma}, \hat{D}_{\gamma} \middle| \{ \Phi(t) \} \right) \propto 
  && P\left( \{ \Phi(t) \} \middle| \bm{c}_{\gamma}, \hat{D}_{\gamma} \right)
  \nonumber \\
  && \times P_{\mathrm{prior}}\left( \bm{c}_{\gamma}, \hat{D}_{\gamma} \right).
  \label{equ:post}
\end{eqnarray}
The Gaussian distribution part in Eqs.~(\ref{equ:prior}) and (\ref{equ:post})
describes the probability of the $\bm{c}_{\gamma}$, which are the coefficients of the basis functions,
while the inverse-gamma distribution part
describes that of the noise intensity $\hat{D}_{\gamma}$.
We define the estimated phase equation by choosing the parameters that maximize the posterior distribution, Eq.~(\ref{equ:post}).

\par
We implement the model selection to determine the maximum order of the Fourier series in each coupling function.
The model complexity depends on $M_{\gamma \nu}$,
which determines the dimensionality of $\bm{c}_{\gamma}$
and the maximum order of the Fourier series,
as shown in Eq.~(\ref{equ:phase_coupling}).
For instance, the coupling function $\hat{\Gamma}_{\gamma \nu}$ is zero if $M_{\gamma \nu}$ = 0,
while it contains higher-order Fourier series if $M_{\gamma \nu}$ is large.
We determine $M_{\gamma \nu}$ by choosing the value of $M'_{\gamma \nu} \in {\bf Z}~(0 \leq M'_{\gamma \nu} \leq M_{\max})$,
which maximizes the model evidence $P_{\mathrm{ME}}\left(\{\Phi(t)\} \middle| M'_{\gamma \nu} \right)$ as follows:
\begin{eqnarray}
  && M_{\gamma \nu} = 
  \argmax_{0 \leq M'_{\gamma \nu} \leq M_{\max}} P_{\mathrm{ME}}\left(\{\Phi(t)\} \middle| M'_{\gamma \nu} \right),
  \\
  && P_{\mathrm{ME}}\left( \{\Phi(t)\} \middle| M'_{\gamma \nu} \right) = 
  \frac{ P\left( \{ \Phi(t) \} \middle| {\bf c}_{\gamma}, \hat{D}_{\gamma} \right)
         P_{\mathrm{prior}}\left( {\bf c}_{\gamma}, \hat{D}_{\gamma} \right) }
       { P_{\mathrm{post}}\left( {\bf c}_{\gamma}, \hat{D}_{\gamma} \middle| \{ \Phi(t) \} \right) }.
  \nonumber
\end{eqnarray}
In this study,
we define $M_{\max} := 15$ and all $M_{\gamma \nu}$ are determined in the range of less than $M_{\max}$.

%% Case1 %%%%%%%%%%%%%%%%%%%%%%%%%%%%%%%
\section{Results}
\label{sec:results}

The following three cases are mainly considered in this section:
two networks of FHN elements (Sec.~\ref{subsec:case1}),
three networks of FHN elements (Sec.~\ref{subsec:case2}), and
two networks of van der Pol oscillators (Sec.~\ref{subsec:case3}).
In each case,
we first explain the dynamics of a system and how to obtain phase time-series from observed time-series.
Second,
we show a representative result of our method.
Finally,
we examine the effect of the choice of observed variables on the result of our method.
As a further study,
we also investigate two networks of globally coupled phase oscillators in partially locked states (Sec.~\ref{subsec:case4}).

\subsection{Case~1: Two networks of FHN elements}
\label{subsec:case1}

\begin{figure*}
  \includegraphics[width=0.9\textwidth]{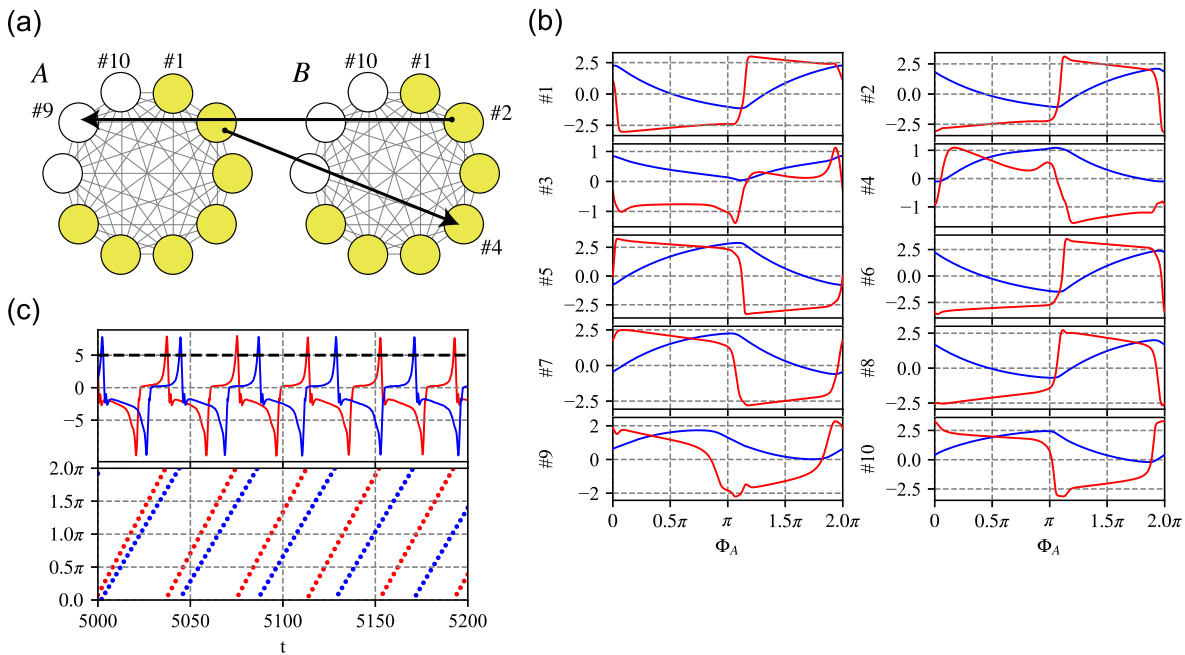}
  \caption{
    Properties of the dynamical system for Case~1.
    {\bf (a)}~
    Schematic diagram of networks $A$ and $B$,
    each of which has 7 excitable elements (yellow circles) and 3 oscillatory elements (white circles).
    An arrow indicates an external coupling between two networks,
    corresponding to Eqs.~(\ref{equ:connection_case1_AB}) and (\ref{equ:connection_case1_BA}).
    {\bf (b)}~
    Waveforms of network $A$.
    Each panel shows $u_i^A$ (blue) and $v_i^A$ (red) of elements $i~(i = 1, 2, \ldots, 10)$
    for one period as a function of $\Phi_A$.
    Network $B$ possesses a similar trajectory due to $K_{i j}^A = K_{i j}^B$.
    {\bf (c)}~
    Time-series of the observed variables and phase variables.
    The time-series of $\sum_i v_i^A(t)$ (red) and $\sum_i v_i^B(t)$ (blue) are displayed in the upper panel.
    The Poincar\'e section (black dashed line) is fixed at
    $\sum_i v_i^A = 5.0$ for network $A$ and
    $\sum_i v_i^B = 5.0$ for network $B$.
    The time-series of $\Phi_A(t)$ (red) and $\Phi_B(t)$ (blue) are displayed in the lower panel.
    The value of the phase is zero
    when the trajectory of each mean field intersects the Poincar\'e section.
    The intensity of the noise, $\eta_i^{\gamma}(t)$, is set to $\sigma = 0.01$.
  }
  \label{fig:result_case1_explain}
\end{figure*}

\begin{figure*}
  \includegraphics[width=0.9\textwidth]{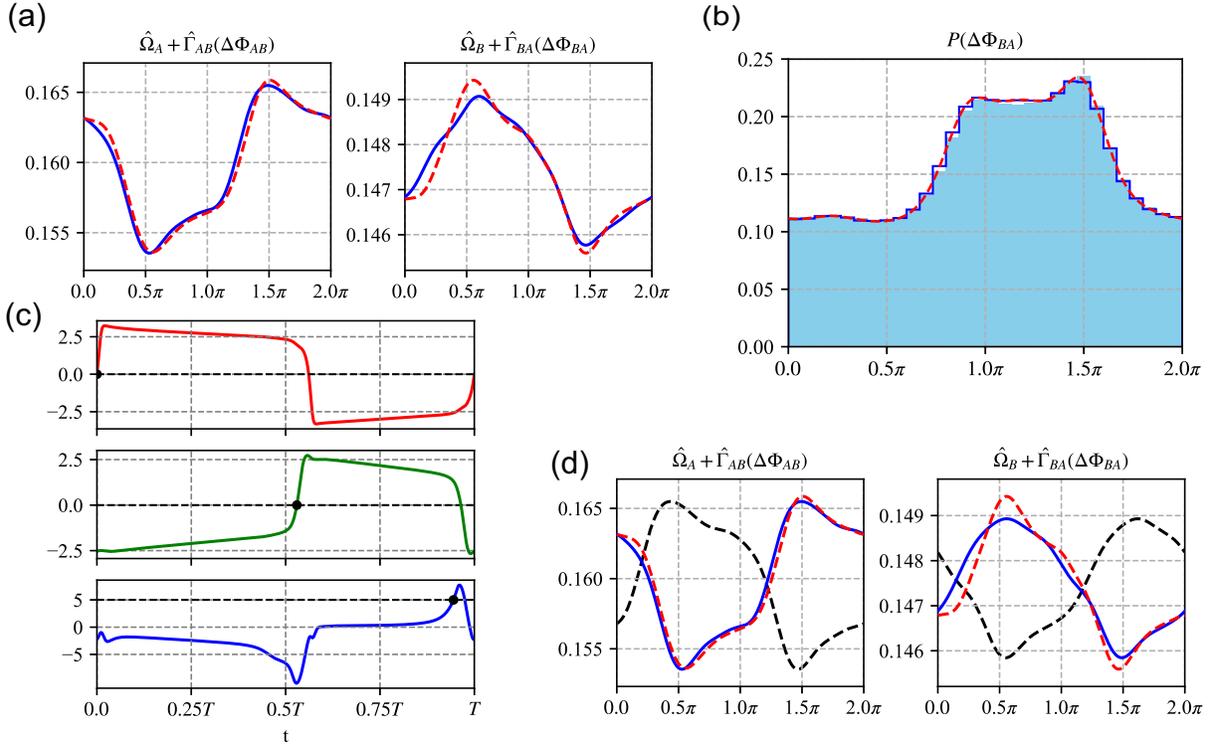}
  \caption{
    A representative result of our method for Case~1.
    {\bf (a)}~
    Phase equations of networks $A$ (left) and $B$ (right) estimated from the time-series of $v_5^A$ and $v_5^B$.
    The length of the time-series satisfies
    $0 \leq |\Delta \Phi_{AB}(t)| \leq 200\pi ~ (t \simeq 5.5 \times 10^4)$.
    The blue solid curves are estimated using the Bayesian approach, whereas
    the red dashed curves are calculated using the numerical method~\cite{nakao_phase_2018}.
    {\bf (b)}~
    Distribution $P(\Delta \Phi_{BA})$ calculated from time-series (cyan) and
    the distribution reproduced from the estimated model (blue).
    Each distribution is obtained from the time-series up to $t = 5.0 \times 10^5$.
    The distribution analytically calculated from the true value of
    $\hat{\Gamma}_{AB}(-\Delta \Phi_{BA}) - \hat{\Gamma}_{BA}(\Delta \Phi_{BA})$ is also plotted (red).
    {\bf (c)}~
    Waveforms of $v_5^A(t)$ (top), $v_8^A(t)$ (middle), and $\sum_i v_i^A(t)$ (bottom) for one period on the limit-cycle orbit.
    The Poincar\'e sections set for each variable are also plotted (black dashed line).
    The phase shift results from the difference in the intersection time.
    {\bf (d)}~
    Phase equations of networks $A$ (left) and $B$ (right) estimated from the time-series of $v_5^A$ and $v_8^B$.
    The red dashed curves are the same as those presented in panel~(a).
    The black dashed curves are estimated using the Bayesian approach, and
    the blue solid curves are obtained by subtracting the phase shift from the originally estimated curves.
  }
  \label{fig:result_case1}
\end{figure*}

\begin{figure*}
  \includegraphics[width=0.9\textwidth]{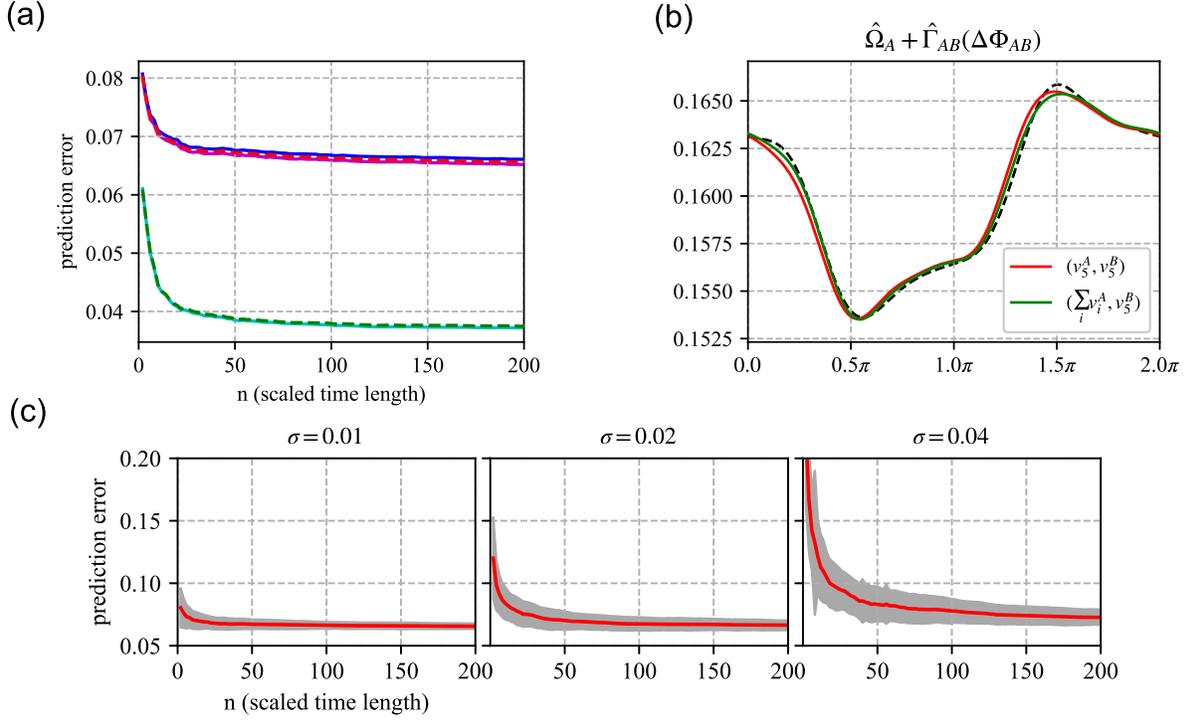}
  \caption{
    Statistics of the prediction error calculated from 100 trials for Case~1.
    {\bf (a)}~
    Mean of the prediction error pertaining to the phase equation of network $A$ calculated from 100 trials.
    The intensity of noise is $\sigma = 0.01$.
    The length of the time-series satisfies
    $0 \leq |\Delta \Phi_{AB}(t)| \leq 2n\pi$,
    where $n$ is the horizontal axis value.
    The error (vertical axis) is calculated by integrating the difference between two curves
    (e.g., the red dashed curve and blue solid curve shown in Fig.~\ref{fig:result_case1}(a))
    from $0$ to $2\pi$ and then divided by true value of
    $\pi ( \max_{\Delta \Phi} \hat{\Gamma}_{AB}(\Delta \Phi) - \min_{\Delta \Phi} \hat{\Gamma}_{AB}(\Delta \Phi) )$.
    The sets of observed variables are
    $(v_5^A, v_5^B)$ (red),
    $(v_5^A, \sum_i v_i^B)$ (blue),
    $(\sum_i v_i^A, v_5^B)$ (green),
    $(\sum_i v_i^A, \sum_i v_i^B)$ (cyan), and
    $(v_5^A, v_8^B)$ (magenta).
    {\bf (b)}~
    Phase equations of network $A$ estimated from a time-series.
    The length of the time-series satisfies $0 \leq |\Delta \Phi_{AB}(t)| \leq 200\pi$.
    The sets of observed variables are
    $(v_5^A, v_5^B)$ (red) and
    $(\sum_i v_i^A, v_5^B)$ (green).
    The black dashed curve is the result of the numerical method~\cite{nakao_phase_2018}.
    {\bf (c)}~
    Mean (red curve) and standard deviation (gray) of prediction error calculated from 100 trials.
    The set of observed variables is $(v_5^A, v_5^B)$.
    The intensity of noise is $\sigma = 0.01$ (left), $0.02$ (middle), and $0.04$ (right).
  }
  \label{fig:result_case1_Stat}
\end{figure*}

We consider a case of two ($N = 2$) FHN networks exhibiting collective oscillations.
Figure~\ref{fig:result_case1_explain}(a) shows that networks $A$ and $B$ interact with each other.
The state variable of each element in network $\gamma$ is represented by
$\bm{X}_i^{\gamma} = (u_i^{\gamma}, v_i^{\gamma})$~($i = 1, 2, \ldots, 10$),
which follows
\begin{eqnarray}
  \dot{u}_i^{\gamma}
  &=& \delta^{\gamma} (a + v_i^{\gamma} - b u_i^{\gamma}),
  \label{equ:u_case1} \\
  \dot{v}_i^{\gamma}
  &=& v_i^{\gamma} - \frac{(v_i^{\gamma})^3}{3} - u_i^{\gamma} + I_i
  + \sum_{j \neq i}^{10} K_{ij}^{\gamma} (v_j^{\gamma} - v_i^{\gamma})
  \nonumber \\ 
  & & + \epsilon \sum_{j = 1}^{10} C_{ij}^{\gamma \nu} (v_j^{\nu} - v_i^{\gamma}) + \eta_i^{\gamma}(t),
  \label{equ:v_case1}
\end{eqnarray}
where $(\gamma, \nu) = (A, B), (B, A)$.
The value of $K_{ij}^{\gamma} \in {\bf R}$ determines
the intensity of the internal coupling from element $j$ to element $i$ in network $\gamma$,
and the set $\{K_{ij}^\gamma\}_{i,j}$ determines the structure of network $\gamma$.
The connection from element $j$ in network $\nu$ to element $i$ in network $\gamma$
is determined by $C_{ij}^{\gamma \nu} \in \{0, 1\}$.
Each element can be oscillatory or excitable depending on the value of the parameter $I_i$.
Each value is $I_i = 0.2$ for $i = 1, \ldots, 7$,
which exhibits excitable dynamics,
and $I_i = 0.8$ for $i = 8, 9, 10$,
which exhibits oscillatory one. 
The time constants are set to $\delta^A = 0.08$ and $\delta^B = 0.073$.
The other parameters are $a = 0.7$ and $b = 0.8$.
The Gaussian noise $\eta_i^{\gamma}(t)$ satisfies
$\langle \eta^{\gamma}_i(t) \rangle = 0$,
$\langle \eta^{\gamma}_i(t) \eta^{\nu}_j(s) \rangle = \sigma^2 \delta_{\gamma \nu} \delta_{ij} \delta(t - s)$.

\par
In this case,
we design the structures of two networks to be the same, i.e., $K_{ij}^A = K_{ij}^B$.
The intensity of the internal coupling, $K_{ij}^A (= K_{ij}^B)$,
is randomly and independently drawn from a uniform distribution $[-0.6,~ 0.6]$.
Note that we choose the parameters so that each network can have collective oscillation which is a unique stable solution.
Figure~\ref{fig:result_case1_explain}(b) shows the waveforms of all elements in network $A$ for one period.
Network $B$ has similar waveforms as network $A$,
but their periods of the limit-cycle solutions are slightly different.
The frequencies of the collective oscillations of the two networks are
$\Omega_A \simeq 0.159$ and $\Omega_B \simeq 0.148$
when there are no perturbations to the two networks.

\par
The connection between two elements belonging to different networks is given as follows:
\begin{eqnarray}
  C_{ij}^{AB} &=& 
  \begin{cases}
    1 & \mathrm{for} \hspace{1mm} (i, j) = (9, 2), \\
    0 & \mathrm{otherwise},
  \end{cases}
  \label{equ:connection_case1_AB} \\
  C_{ij}^{BA} &=& 
  \begin{cases}
    1 & \mathrm{for} \hspace{1mm} (i, j) = (4, 2), \\
    0 & \mathrm{otherwise}.
  \end{cases}
  \label{equ:connection_case1_BA}
\end{eqnarray}
Figure~\ref{fig:result_case1_explain}(a) shows the schematic diagram of the connection between the two networks.
The intensity of the external couplings is $\epsilon = 0.01$ and so weak that the collective oscillation of each network persists.

\par
To apply the Bayesian inference,
we generate a phase sampling data, $\{\Phi(t)\}$, from a time-series of an observed variable,
which is chosen only one for each network.
We choose $v_5^A$ or $\sum_i v_i^A$ for network $A$, and
$v_5^B$, $v_8^B$, or $\sum_i v_i^B$ for network $B$
as the observed variables.
The $5$th element is excitable, and the $8$th element is oscillatory.
We record the times $T_k^{\gamma}$
when the network $\gamma$ intersects the Poincar\'e section at the $k$th time.
The Poincar\'e section is set for each observed variable.
In this case,
we select $v_5^A = 0.0$ or $\sum_i v_i^A = 5.0$ for network $A$,
and $v_5^B = 0.0$, $v_8^B = 0.0$, or $\sum_i v_i^B = 5.0$ for network $B$.
These sections are chosen so that the trajectory of each network can intersect only once for one period.

\par
To obtain $\{\Phi(t)\}$ using the train of $T^{\gamma}_k$, 
we interpolate the phase $\Phi_{\gamma}(t_l) \in [0, 2\pi)$
at time $t_l = l \Delta t$~$(l = 0, 1, \ldots, L)$.
The value of phase $\Phi_{\gamma}(t_l)$,
where $t_l$ is between $T_{k}^{\gamma}$ and $T_{k+1}^{\gamma}$,
is obtained as follows:
\begin{eqnarray}
  \Phi_{\gamma}(t_l) = 2\pi \frac{t_l - T^{\gamma}_{k}}{T^{\gamma}_{k+1} - T^{\gamma}_{k}}
  ~( T^{\gamma}_{k} \leq t_l < T^{\gamma}_{k+1} ),
  \label{equ:interpolation}
\end{eqnarray}
where $T_{k}^{\gamma}$ satisfies $\Phi_{\gamma}(T_{k}^{\gamma}) = 0$.
The sampling interval is $\Delta t = 2.0$.
Figure~\ref{fig:result_case1_explain}(c)
shows the phase time-series of networks $A$ and $B$
generated from the time-series of $\sum_i v_i^A$ and $\sum_i v_i^B$.

\par
We first confirm whether our method succeeds or not,
and then investigate how the choice of observed variables affects the result of our method.
Here,
the hyperparameters in Eq.~(\ref{equ:prior}) are set to
$\alpha_0 = \beta_0 = \lambda_0 = 1.0 \times 10^{-3}$,
and the noise intensity is $\sigma = 0.01$.
Figure~\ref{fig:result_case1}(a) shows the phase equations of networks $A$ and $B$
estimated from the time-series of $v_5^A$ and $v_5^B$.
The length of the time-series satisfies
$0 \leq |\Delta \Phi_{AB}(t)| \leq 200\pi ~(t \simeq 5.5 \times 10^4)$.
The estimated phase equations are similar to the result obtained from the numerical method~\cite{nakao_phase_2018}.
Figure~\ref{fig:result_case1}(b) shows that
the estimated model shown in Fig.~\ref{fig:result_case1}(a) reproduced
the distribution $P(\Delta \Phi_{BA})$ calculated from the phase time-series up to $t = 5.0 \times 10^5$.

\par
Note that the phase value shifts when the observed variable changes,
because the time to intersect the Poincar\'e section also varies
depending on the trajectory of the observed variable.
Figure~\ref{fig:result_case1}(c) indicates that
the time corresponding to $\Phi_A(t) = 0$ varies depending on the observed variable.
Since the phase coupling function is the function of the phase difference,
the $\hat{\Gamma}_{AB}$ and $\hat{\Gamma}_{BA}$ also shift depending on the phase shift,
as shown in Fig.~\ref{fig:result_case1}(d),
where the phase equations estimated from the time-series of $v_5^A$ and $v_8^B$, instead of $v_5^A$ and $v_5^B$, are plotted.
In this study,
the phase shift is subtracted from the estimated model.

\par
We calculate the prediction error of 100 trials for each set of observed variables
to investigate the effect of the choice of observed variables on our method.
The prediction error is calculated by comparing between the two phase equations,
one of which is obtained using our Bayesian method
and the other is calculated from the numerical method~\cite{nakao_phase_2018}.
Figure~\ref{fig:result_case1_Stat}(a) shows
the comparison of the mean error pertaining to the phase equation of network $A$ between the sets of observed variables
$(v_5^A, v_5^B)$,
$(v_5^A, \sum_i v_i^B)$,
$(\sum_i v_i^A, v_5^B)$,
$(\sum_i v_i^A, \sum_i v_i^B)$, and
$(v_5^A, v_8^B)$,
with the noise intensity, $\sigma = 0.01$.
The observed variables have several properties:
$v_5^A$ and $v_5^B$ are observed from one of the excitable elements in networks $A$ and $B$, respectively;
$v_8^B$ is observed from one of the oscillatory elements in network $B$; and
$\sum_i v_i^A$ and $\sum_i v_i^B$ are the mean-field of all the elements in networks $A$ and $B$, respectively.
Each mean-field includes both excitable and oscillatory elements.
The horizontal axis value, $n$, represents the length of phase time-series,
e.g., $0 \leq | \Delta \Phi_{AB}(t) | \leq 2n\pi$.
When the value of $| \Delta \Phi_{AB}(t) |$ increases by $2\pi$,
the phase time-series covers all of the state space of $\hat{\Gamma}_{AB}(\Delta \Phi_{AB})$.
The vertical axis value represents the error value,
which is calculated by integrating the difference between the estimated phase equation and the numerically calculated one.
This value is normalized by the amplitude of $\hat{\Gamma}_{AB}$,
that is,
the value of error is divided by 
$\pi ( \max_{\Delta \Phi} \hat{\Gamma}_{AB}(\Delta \Phi) - \min_{\Delta \Phi} \hat{\Gamma}_{AB}(\Delta \Phi) )$.
We can find that the error decreases with an increase in the number of data points,
and this decreasing trend is similar regardless of the types of observed variables.
Figure~\ref{fig:result_case1_Stat}(b)
shows two estimated phase equations of network $A$;
one is for the set of observed variables $(v_5^A, v_5^B)$ and
the other is for $(\sum_i v_i^A, v_5^B)$.
The length of the time-series satisfies
$0 \leq |\Delta \Phi_{AB}(t)| \leq 200\pi$.
The two curves are similar,
and this fact is consistent with the results shown in Fig.~\ref{fig:result_case1_Stat}(a).
Figures~\ref{fig:result_case1_Stat}(a) and \ref{fig:result_case1_Stat}(b) indicate that
similar phase coupling functions are extracted by our method regardless of the types of observed variables.
Figure~\ref{fig:result_case1_Stat}(c)
shows the mean and standard deviation calculated from 100 trials
with the noise intensity $\sigma = 0.01, 0.02, 0.04$
for the set of observed variables $(v_5^A, v_5^B)$.
Although the standard deviation of error increases with $\sigma$ becomes large,
the convergence property is maintained
as long as the noise is so weak that the collective oscillation persists.

%% Case2 %%%%%%%%%%%%%%%%%%%%%%%%%%%%%%%
\subsection{Case~2: Three networks of FHN elements}
\label{subsec:case2}

\begin{figure*}
  \includegraphics[width=0.9\textwidth]{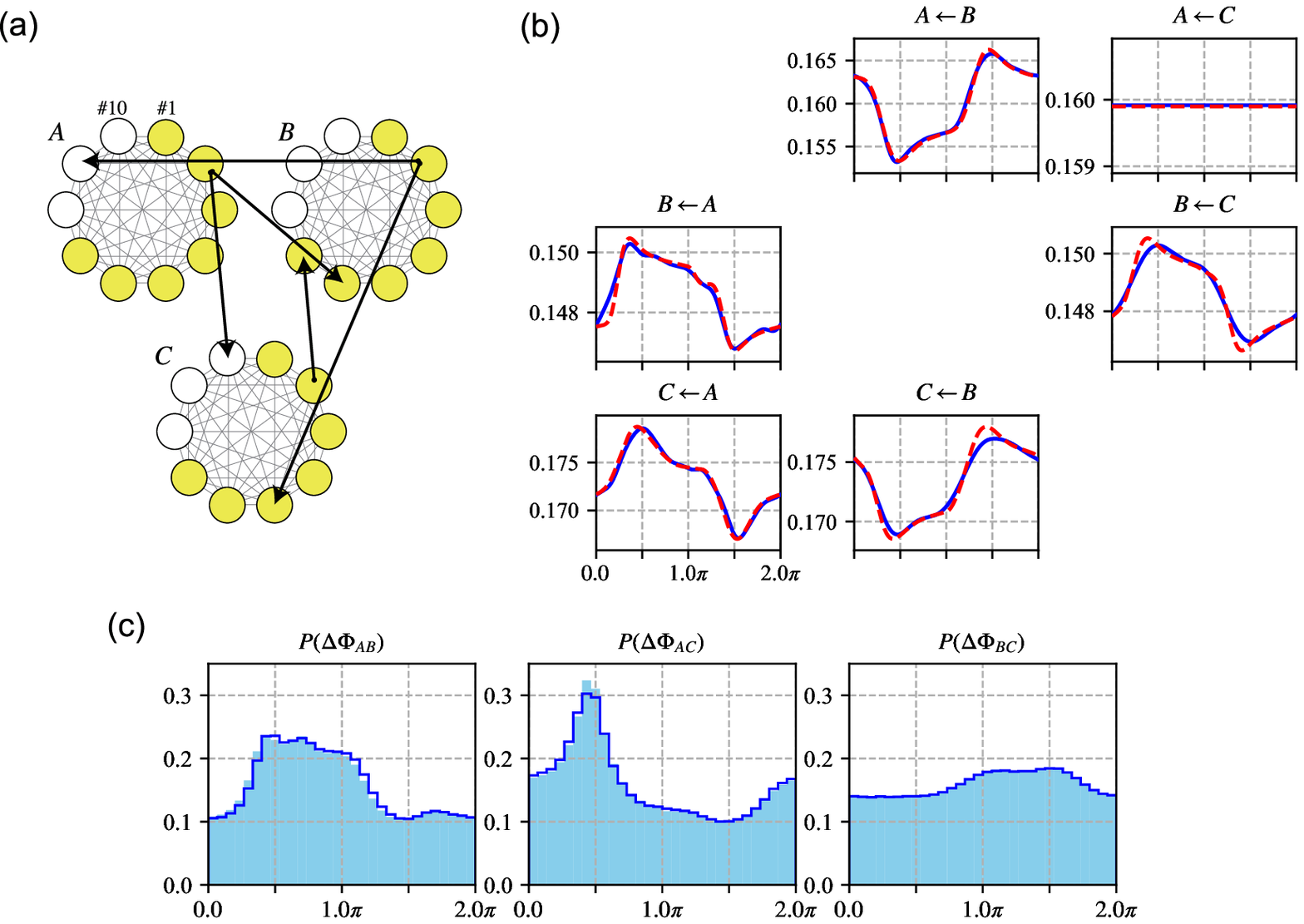}
  \caption{
    Property of the dynamical system and
    a representative result of the model identification for  Case~2.
    {\bf (a)}~
    Schematic diagram of networks $A$, $B$, and $C$,
    each of which has 7 excitable elements (yellow circles) and 3 oscillatory elements (white circles).
    An arrow indicates an external coupling between networks,
    which corresponds to Eq.~(\ref{equ:connection_case2}).
    {\bf (b)}~
    Phase coupling functions with constant term, $\hat{\Omega}_{\gamma}$,
    estimated from the time-series of $\sum_i v_i^A$, $\sum_i v_i^B$, and $\sum_i v_i^C$.
    The length of the time-series satisfies
    $0 \leq |\Delta \Phi_{AB}(t)| \leq 200\pi ~ (t \simeq 9.4 \times 10^4)$.
    The blue solid curves are estimated using the Bayesian approach,
    whereas the red dashed curves are calculated using the numerical method.
    The sender and receiver indexes of the coupling functions are shown at the top of each panel.
    {\bf (c)}~
    Distributions $P(\Delta \Phi_{AB})$ (left), $P(\Delta \Phi_{AC})$ (middle), and $P(\Delta \Phi_{BC})$ (right)
    calculated from time-series (cyan) and
    the distributions reproduced using the estimated model (blue).
    Each distribution is obtained from the time-series up to $t = 5.0 \times 10^5$.
  }
  \label{fig:Result_case2}
\end{figure*}

\begin{figure*}
  \includegraphics[width=0.9\textwidth]{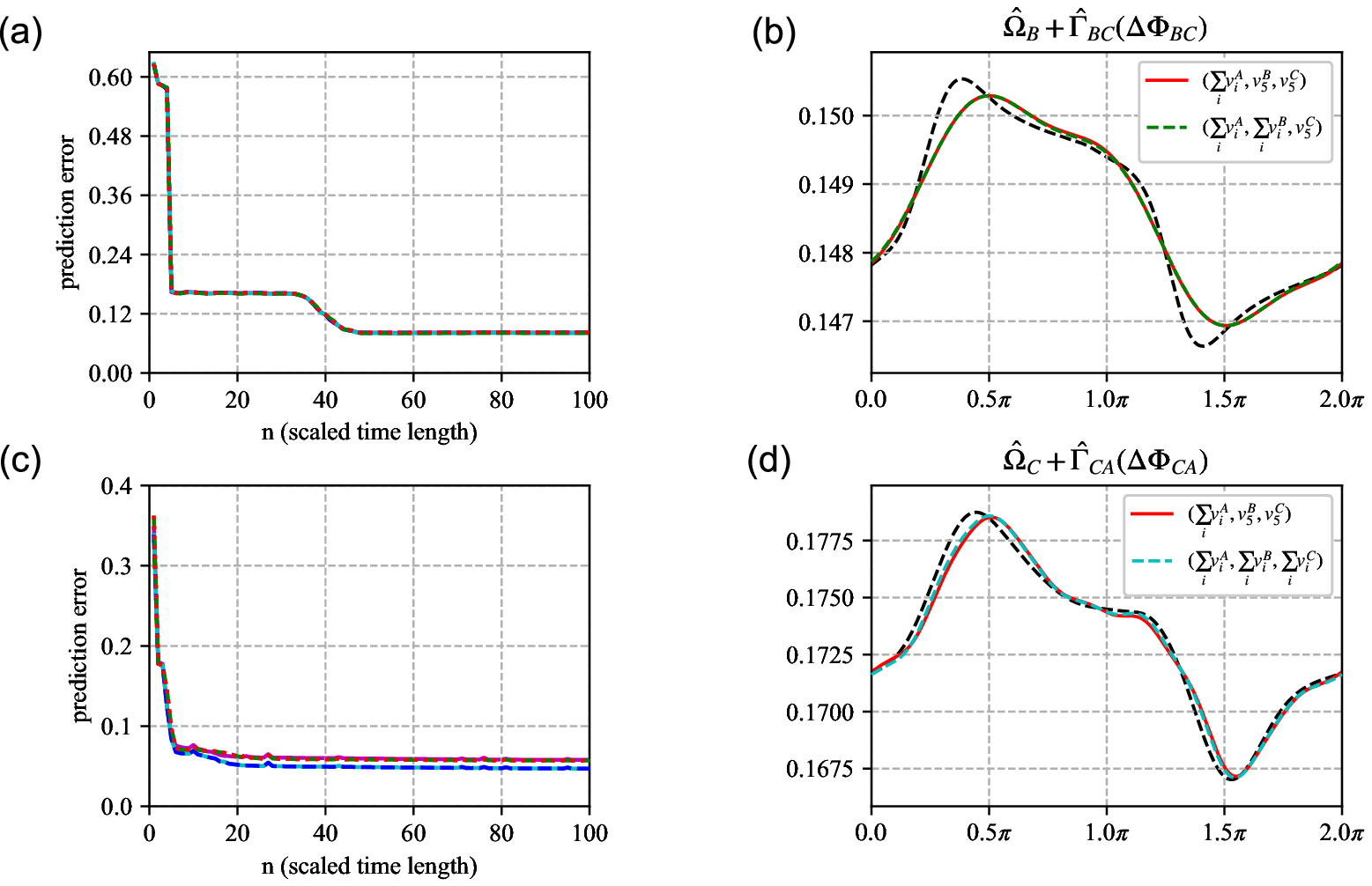}
  \caption{
    Statistics of the prediction error calculated from 100 trials for Case~2.
    {\bf (a)}~
    Mean of the prediction error pertaining to the phase coupling function,
    $\hat{\Gamma}_{BC}(\Delta \Phi_{BC})$,
    calculated from 100 trials.
    The length of the time-series satisfies
    $0 \leq |\Delta \Phi_{BC}(t)| \leq 2n\pi$,
    where $n$ is the horizontal axis value.
    The value of error (vertical axis) is divided by
    $\pi ( \max_{\Delta \Phi} \hat{\Gamma}_{BC}(\Delta \Phi) - \min_{\Delta \Phi} \hat{\Gamma}_{BC}(\Delta \Phi) )$.
    The sets of observed variables are
    $(\sum_i v_i^A, v_5^B, v_5^C)$ (red),
    $(\sum_i v_i^A, v_5^B, \sum_i v_i^C)$ (blue),
    $(\sum_i v_i^A, \sum_i v_i^B, v_5^C)$ (green),
    $(\sum_i v_i^A, \sum_i v_i^B, \sum_i v_i^C)$ (cyan), and
    $(\sum_i v_i^A, v_5^B, v_8^C)$ (magenta).
    These five curves have a similar shape to each other.
    {\bf (b)}~
    Phase coupling functions $\hat{\Gamma}_{BC}(\Delta \Phi_{BC})$ estimated from a time-series.
    The length of the time-series satisfies $0 \leq |\Delta \Phi_{AB}(t)| \leq 200\pi$.
    The sets of observed variables are
    $(\sum_i v_i^A, v_5^B, v_5^C)$ (red) and
    $(\sum_i v_i^A, \sum_i v_i^B, v_5^C)$ (green).
    The black dashed curve is the result of the numerical method.
    {\bf (c) \& (d)}~
    Results for $\hat{\Gamma}_{CA}(\Delta \Phi_{CA})$ are shown in the same way as panels~(a) and (b).
  }
  \label{fig:Result_case2_Stat}
\end{figure*}

We also investigated a case of three ($N = 3$) FHN networks.
Figure~\ref{fig:Result_case2}(a) shows the schematic diagram of the three networks.
This case is a generalization of Case~1 in terms of the number of networks and the network structure.
The dynamics of element $i~(i = 1, 2, \ldots, 10)$ in network $\gamma \in \{A, B, C\}$ is described as follows:
\begin{eqnarray}
  \dot{u}_i^{\gamma}
  &=& \delta^{\gamma} (a + v_i^{\gamma} - b u_i^{\gamma}),
  \label{equ:u_case2} \\
  \dot{v}_i^{\gamma}
  &=& v_i^{\gamma} - \frac{(v_i^{\gamma})^3}{3} - u_i^{\gamma} + I_i
  + \sum_{j \neq i}^{10} K_{ij}^{\gamma} (v_j^{\gamma} - v_i^{\gamma})
  \nonumber \\ 
  & & + \epsilon \sum_{\nu \neq \gamma} \sum_{j=1}^{10} C_{ij}^{\gamma \nu} (v_j^{\nu} - v_i^{\gamma})
  + \eta_i^{\gamma}(t),
  \label{equ:v_case2}
\end{eqnarray}
where $\nu \in \{A, B, C\}$ is also index of networks, which is different from $\gamma$,
i.e., $\nu \neq \gamma$.
The values of $I_i$, $a$, and $b$ are the same as those in Sec.~\ref{subsec:case1}.
The time constants are set to $\delta^A = 0.080$, $\delta^B = 0.096$, and $\delta^C = 0.086$.
The noise intensity is set to $\sigma = 0.005$.

\par
We design each network structure to be different,
i.e., $K_{ij}^A \neq K_{ij}^B \neq K_{ij}^C$,
each of which is randomly and independently drawn from a uniform distribution $[-0.6,~0.6]$.
The waveforms of the limit-cycle solution of each network are different,
mainly due to the differences in network structures
(Fig.~\ref{fig:Append_case2} in Appendix~\ref{sec:appendix}).
Figure~\ref{fig:Result_case2}(a) shows that
the connection between networks is given by
\begin{eqnarray}
  \begin{cases}
    C_{ 9~2}^{A B} = 
    C_{ 6~2}^{B A} = 
    C_{10~2}^{C A} = 
    C_{ 7~2}^{B C} = 
    C_{ 5~2}^{C B} = 1,
    \\
    C_{i j}^{\gamma \nu} = 0 ~~~ \text{otherwise},
  \end{cases}
  \label{equ:connection_case2}
\end{eqnarray}
where there are no couplings from network $C$ to network $A$,
thus,
the phase coupling function, $\hat{\Gamma}_{AC}$, should be zero.
The intensity of the external couplings is set to $\epsilon = 0.01$.

\par
In this case,
we choose $\sum_i v_i^A$ for network $A$,
$v_5^B$ or $\sum_i v_i^B$ for network $B$, and
$v_5^C$, $v_8^C$, or $\sum_i v_i^C$ for network $C$ as the observed variables.
The Poincar\'e section is set to $\sum_i v_i^A = 0.0$ for network $A$,
$v_5^B = 0.0$ or $\sum_i v_i^B = -5.0$ for network $B$, and
$v_5^C = 0.0$, $v_8^C = 0.0$, or $\sum_i v_i^C = 2.0$ for network $C$.
The hyperparameters in Eq.~(\ref{equ:prior}) are set to
$\alpha_0 = \beta_0 = \lambda_0 = 1.0 \times 10^{-2}$.

\par
Figure~\ref{fig:Result_case2}(b) is a representative result of our method.
This result is obtained from the time-series of $\sum_i v_i^A$, $\sum_i v_i^B$, and $\sum_i v_i^C$,
which satisfy $0 \leq |\Delta \Phi_{AB}(t)| \leq 200\pi ~ (t \simeq 9.4 \times 10^4)$.
Figure~\ref{fig:Result_case2}(c) indicates that the estimated model shown in Fig.~\ref{fig:Result_case2}(b)
can reproduce similar distributions $P(\Delta \Phi_{AB})$, $P(\Delta \Phi_{AC})$, and $P(\Delta \Phi_{BC})$,
which are calculated from the time-series up to $t = 5.0 \times 10^5$.

\par
Further,
we also investigated the prediction error for each observed variable in the same way as Sec.~\ref{subsec:case1}.
Figure~\ref{fig:Result_case2_Stat}(a) shows
the mean value of the prediction error pertaining to the phase coupling function, $\hat{\Gamma}_{BC}$, calculated from 100 trials,
and the values of error are similar regardless of the observed variables.
Figure~\ref{fig:Result_case2_Stat}(b) shows the two estimations of $\hat{\Gamma}_{BC}$;
one is for the set of observed variables $(\sum_i v_i^A, v_5^B, v_5^C)$
and the other is for $(\sum_i v_i^A, \sum_i v_i^B, v_5^C)$,
and these two curves are similar to each other.
Figures~\ref{fig:Result_case2_Stat}(c) and \ref{fig:Result_case2_Stat}(d)
show the results for $\hat{\Gamma}_{CA}$ in the same way as
Figs.~\ref{fig:Result_case2_Stat}(a) and \ref{fig:Result_case2_Stat}(b).

%% Case3 %%%%%%%%%%%%%%%%%%%%%%%%%%%%%%%
\subsection{Case~3: Two networks of van der Pol oscillators}
\label{subsec:case3}

\begin{figure*}
  \includegraphics[width=0.9\textwidth]{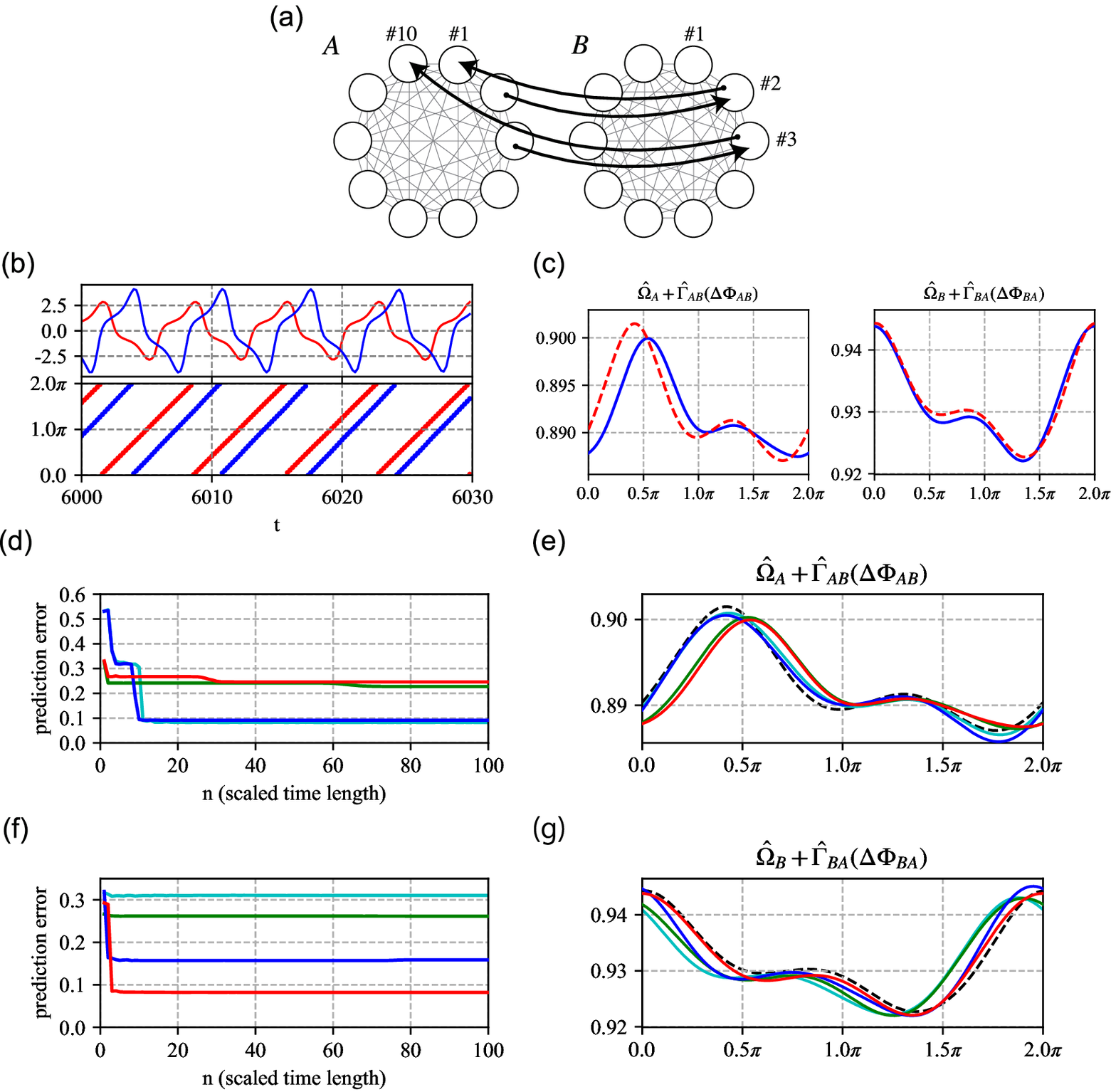}
  \caption{
    Properties of the dynamical system and overall result of the model identification for Case~3.
    {\bf (a)}~
    Schematic diagram of networks $A$ and $B$,
    each of which has 10 oscillatory elements (white circles).
    An arrow indicates an external coupling between two networks,
    corresponding to Eqs.~(\ref{equ:connection_case3_AB}) and (\ref{equ:connection_case3_BA}).
    {\bf (b)}~
    Time-series of the observed variables and phase variables.
    The time-series of $v_5^A $ (red) and $v_2^B $ (blue) are shown in the upper panel, and
    those of $\Phi_A$ (red) and $\Phi_B$ (blue) are shown in the lower panel.
    The sampling interval is set to $\Delta t = 0.2$.
    {\bf (c)}~
    Phase equations of networks $A$ (left) and $B$ (right)
    estimated from the time-series of $v_5^A$ and $v_2^B$.
    The length of the time-series satisfies
    $0 \leq |\Delta \Phi_{AB}(t)| \leq 200\pi$.
    The blue solid curves are estimated using the Bayesian approach, whereas
    the red dashed curves are calculated using the numerical method.
    {\bf (d)}~
    Mean of the prediction error pertaining to the phase equation of network $A$ calculated from 100 trials.
    The length of the time-series satisfies
    $0 \leq |\Delta \Phi_{AB}(t)| \leq 2n\pi$,
    where $n$ is the value of the horizontal axis.
    The value of error (vertical axis) is divided by the true value of
    $\pi ( \max_{\Delta \Phi} \hat{\Gamma}_{AB}(\Delta \Phi) - \min_{\Delta \Phi} \hat{\Gamma}_{AB}(\Delta \Phi) )$.
    The sets of observed variables are
    $(v_5^A, v_2^B)$ (red),
    $(v_{10}^A, v_2^B)$ (blue),
    $(v_5^A, v_{10}^B)$ (green), and
    $(v_{10}^A, v_{10}^B)$ (cyan).
    {\bf (e)}~
    Phase equations of network $A$ estimated from each set of observed variables.
    The length of the time-series satisfies $0 \leq |\Delta \Phi_{AB}(t)| \leq 200\pi$.
    The colors of curves and the sets of observed variables are the same as those in panel~(d).
    The black dashed curve is the result of the numerical method.
    {\bf (f) \& (g)}~
    Results for network $B$ are shown in the same way as panels~(d) and (e).
  }
  \label{fig:Result_case3}
\end{figure*}

We also investigated a case of two ($N = 2$) van der Pol networks.
Figure~\ref{fig:Result_case3}(a) shows the schematic diagram of the two networks.
In this case,
we employ another type of element that oscillates slowly (not a fast-slow system such as the FHN element),
and apply the Hilbert transform~\cite{pikovsky_synchronization_2001}
to transform observed time-series into phase time-series,
instead of linear interpolation described in Eq.~(\ref{equ:interpolation}).
The linear interpolation method averages the effect of perturbations to a network on the phase variable over one period,
whereas the Hilbert transform reflects perturbations in phase time-series without averaging.

\par
The state variables of each element in network $\gamma$ is represented by 
$\bm{X}_i^{\gamma} = (u_i^{\gamma}, v_i^{\gamma}) ~ (i = 1, 2, \ldots, 10)$,
which follows
\begin{eqnarray}
  \dot{u}_i^{\gamma}
  &=& v_i^{\gamma},
  \label{equ:u_case3} \\
  \dot{v}_i^{\gamma}
  &=& -u_i^{\gamma} - a^{\gamma} v_i^{\gamma} \bigl( (u_i^{\gamma})^2 - 1 \bigr)
  + \sum_{j \neq i}^{10} K_{i j}^{\gamma} (v_j^{\gamma}-v_i^{\gamma})
  \nonumber \\
  & & + \epsilon \sum_{j=1}^{10}
  C_{i j}^{\gamma \nu} \bigl( (v_j^{\nu})^2 u_i^{\gamma} - (v_i^{\gamma})^2 u_j^{\nu} \bigr)
  + \eta_i^{\gamma}(t),
  \label{equ:v_case3}
\end{eqnarray}
where $(\gamma, \nu) = (A, B), (B, A)$.
The nonlinearity parameters are set to $a^A = 0.5$ and $a^B = 0.3$,
and the noise intensity is $\sigma = 0.005$.
%% K
The structures of two networks are designed to be different, i.e., $K_{ij}^A \neq K_{ij}^B$,
each of which is randomly and independently drawn from a uniform distribution $[-0.6,~0.6]$.
We choose these parameters so that each network can exhibit the collective oscillation
(Fig.~\ref{fig:Append_case3} in Appendix~\ref{sec:appendix}).
Figure~\ref{fig:Result_case3}(a) show that
the connection between two networks is given as follows:
\begin{eqnarray}
  \label{equ:connection_case3_AB}
  C_{ij}^{AB} &=& 
  \begin{cases}
    1 & \mathrm{for} \hspace{1mm} (i, j) = (1, 2), (10, 3), \\
    0 & \mathrm{otherwise},
  \end{cases} \\
  \label{equ:connection_case3_BA}
  C_{ij}^{BA} &=& 
  \begin{cases}
    1 & \mathrm{for} \hspace{1mm} (i, j) = (2, 2), ( 3, 3), \\
    0 & \mathrm{otherwise}.
  \end{cases}
\end{eqnarray}
The intensity of the external couplings is set to $\epsilon = 0.005$.

\par
Here, we transform the observed time-series to the phase time-series using the Hilbert transformation,
which generates a protophase time-series.
We observe the time-series of $v_5^A$ or $v_{10}^A$ from network $A$, and
$v_2^B$ or $v_{10}^B$ from network $B$.
Each time-series is observed from one of the oscillators.
We denote the dynamics observed from network $\gamma$ at time $t$ as $s_{\gamma}(t)$ for simplicity.
To transform $s_{\gamma}(t)$ into protophase time-series $\Theta_{\gamma}(t)$,
we use Hilbert transformation $s_{\gamma}^{\mathcal{H}}(t)$ as follows~\cite{pikovsky_synchronization_2001}:
\begin{eqnarray}
  A_{\gamma}(t) \e^{\mathrm{i} \Theta_{\gamma}(t)}
  = s_{\gamma}(t) + \mathrm{i} s_{\gamma}^{\mathcal{H}}(t),
\end{eqnarray}
where $A_{\gamma}(t)$ is the amplitude of the equation's right-hand side.
Note that protophase $\Theta_{\gamma}(t)$ obtained by Hilbert transformation
of the time-series observed from a nonlinear oscillator
does not increase linearly with time.
Therefore,
we construct a phase time-series, $\Phi_{\gamma}(t)$, from the protophase time-series, $\Theta_{\gamma}(t)$,
as follows~\cite{kralemann_reconstructing_2014, kralemann_reconstructing_2011, kralemann_phase_2008, kralemann_uncovering_2007}:
\begin{eqnarray}
  \Phi_{\gamma}(\Theta_{\gamma})
  = 2\pi \int_0^{\Theta_{\gamma}} f_{\gamma}(\Theta') d\Theta',
\end{eqnarray}
where $f_{\gamma}(\Theta_{\gamma})$ is the probability density function of $\Theta_{\gamma}$.
Figure~\ref{fig:Result_case3}(b) shows the time-series of $v_5^A$, $v_2^B$, $\Phi_A$, and $\Phi_B$,
where each phase increases linearly with time.

\par
Figure~\ref{fig:Result_case3}(c) shows the dynamics of networks $A$ and $B$ estimated from the time-series of $v_5^A$ and $v_2^B$.
The hyperparameters in Eq.~(\ref{equ:prior}) are set to $\alpha_0 = \beta_0 = \lambda_0 = 1.0\times 10^{-3}$,
and the length of the time-series used for the model identification satisfies $0 \leq | \Delta \Phi_{AB}(t)| \leq 200\pi$.

\par
We investigated the effect of the choice of observed variables
in the same way as Secs.~\ref{subsec:case1} and \ref{subsec:case2}.
Figure~\ref{fig:Result_case3}(d) shows the mean of the prediction error pertaining to the phase equation for network $A$.
Figure~\ref{fig:Result_case3}(e) shows the estimated phase equation for network $A$ for each set of observed variables.
Figures~\ref{fig:Result_case3}(f) and \ref{fig:Result_case3}(g)
show the results for network $B$ in the same way as
Figs.~\ref{fig:Result_case3}(d) and \ref{fig:Result_case3}(e).

%% Case4 %%%%%%%%%%%%%%%%%%%%%%%%%%%%%%%
\subsection{Case~4: Partially locked state}
\label{subsec:case4}

\begin{figure*}
  \includegraphics[width=0.9\textwidth]{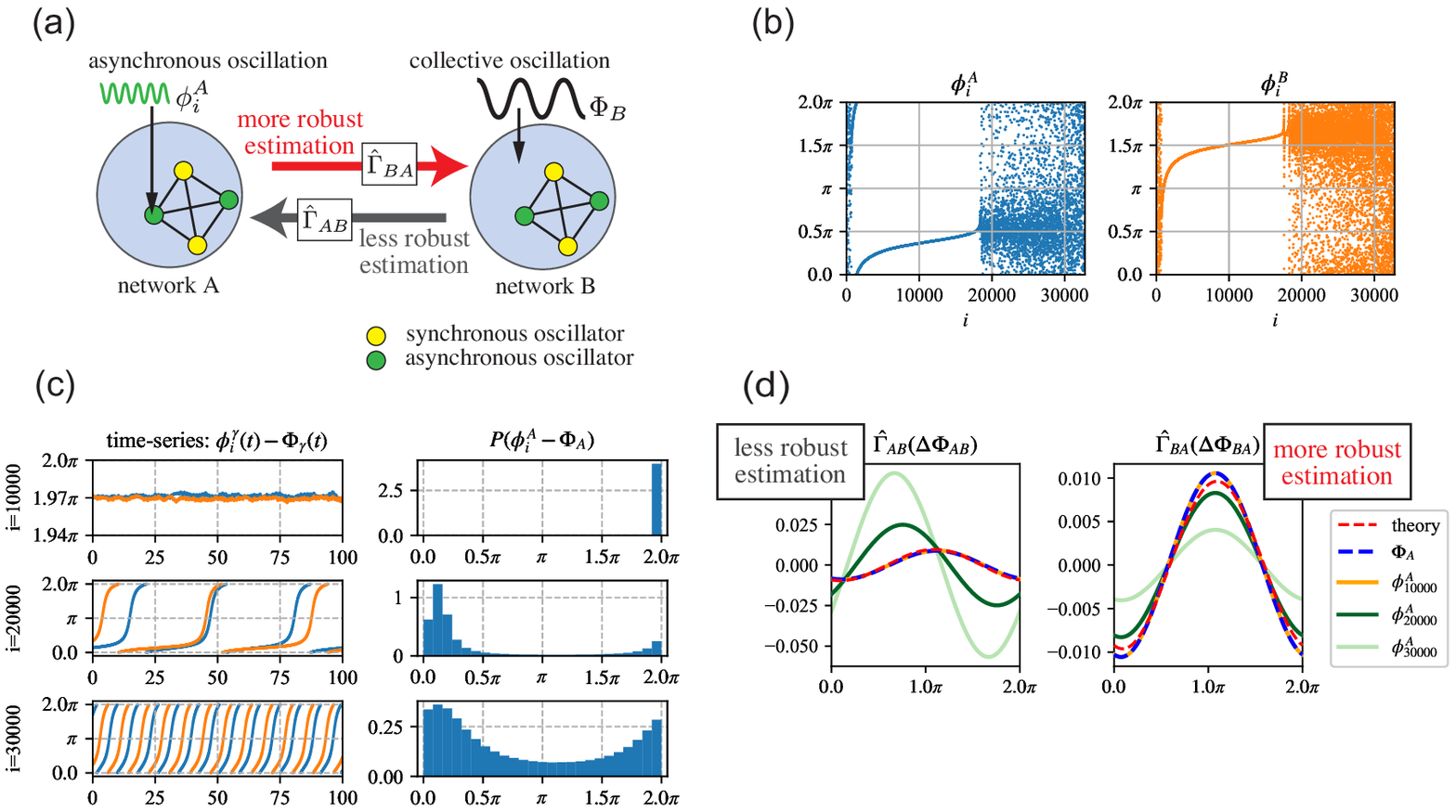}
  \caption{
    Properties of the dynamical system and representative results of model identification for Case~4.
    {\bf (a)}~
    Schematic diagram of networks $A$ and $B$ (blue circles),
    each of which contains synchronous oscillators (yellow dots) and asynchronous oscillators (green dots).
    This diagram illustrates a situation
    where an asynchronous oscillator in network $A$
    and the macroscopic collective oscillation of network $B$ are observed.
    Our study shows that
    the macroscopic phase coupling function from network $A$ to network $B$ (red arrow)
    can be extracted using model identification
    more robustly than in the opposite direction.
    {\bf (b)}~
    Snapshots of the phase oscillators in networks $A$ (left) and $B$ (right).
    The oscillator index $i$ is sorted in increasing order of the natural frequencies.
    {\bf (c)}~
    Time-series and distributions of the phase difference $\phi_i^{\gamma} - \Phi_{\gamma}$.
    The left column is the time-series of $\phi_i^{\gamma}(t) - \Phi_{\gamma}(t)$ of networks $A$ (blue) and $B$ (orange).
    The right column is the distribution $P(\phi_i^A - \Phi_A)$ obtained from the time-series up to $t = 1.0 \times 10^4$.
    Each row corresponds to $i = 10000, 20000, 30000$.
    The distribution $P(\phi_i^B - \Phi_B)$, which is not displayed, has a similar shape to that of network $A$.
    {\bf (d)}~
    Macroscopic phase coupling functions $\hat{\Gamma}_{AB}$ (left) and $\hat{\Gamma}_{BA}$ (right)
    estimated from the time-series of observed variables.
    The correspondence between the colors of lines and the observed variables of network $A$ is shown in the legend.
    The observed variable of network $B$ is the collective phase $\Phi_B$.
    The results are obtained from the time-series up to $t = 1.0 \times 10^4$.
    The red dashed lines are theoretical curves.
  }
  \label{fig:Result_case4}
\end{figure*}

In Case~1 to Case~3, we focused on networks, in which all the elements were in the fully locked state.
As a further study,
we investigated a case of two ($N = 2$) networks of globally coupled phase oscillators in the partially locked state.
Figure~\ref{fig:Result_case4}(a) shows a schematic diagram of two networks (blue circles),
each of which contains synchronous oscillators (yellow dots) and asynchronous oscillators (green dots).

\par
In this subsection,
we discuss which oscillator should be chosen to robustly extract the macroscopic phase coupling function.
From this viewpoint,
we investigated the possibility of extracting the macroscopic phase coupling function
using the time-series of a synchronous, asynchronous oscillator, or macroscopic collective oscillation.
Figure~\ref{fig:Result_case4}(a) illustrates a case
where an asynchronous oscillator in network $A$ and the macroscopic collective oscillation of network $B$ are observed.
To provide our conclusion implied in the schematic diagram,
we will return to Fig.~\ref{fig:Result_case4}(a) at the end of this subsection.

\par
We consider two weakly coupled networks of globally coupled phase oscillators.
The phase of each oscillator in network $\gamma$ is represented by $\phi_i^{\gamma}~(i = 1, 2, \ldots, N_{\gamma})$,
which follows
\begin{eqnarray}
  \dot{\phi}_i^{\gamma}(t)
  &=& \omega_i^{\gamma} + \frac{1}{N_{\gamma}} \sum_{j=1}^{N_{\gamma}} \sin(\phi_j^{\gamma} - \phi_i^{\gamma} - \alpha)
  \nonumber \\
  & & + \frac{\epsilon}{N_{\nu}} \sum_{j=1}^{N_{\nu}} \sin(\phi_j^{\nu} - \phi_i^{\gamma} - \beta),
  \label{equ:dynamics_case4}
\end{eqnarray}
where $(\gamma, \nu) = (A, B)$ or $(B, A)$.
The second term on the right-hand side represents internal coupling between oscillators within the same group,
and the last term represents external coupling between oscillators that belong to different groups.
These couplings are characterized by the coupling phase shifts $\alpha$ and $\beta$.
The intensity of the external coupling is given by $\epsilon \ll 1$.
The natural frequency $\omega_j^{\gamma}$ is drawn from the Lorentzian distribution
with central value $\omega_0^{\gamma}$ and dispersion $w$,
\begin{eqnarray}
  g_{\gamma}(\omega_i^{\gamma})
  = \frac{w}{\pi} \frac{1}{(\omega_i^{\gamma} - \omega_0^{\gamma})^2 + w^2}.
\end{eqnarray}
We introduce the order parameter with modulus $R_{\gamma}(t)$ and collective phase $\Phi_{\gamma}(t)$ as
\begin{eqnarray}
  R_{\gamma}(t) \e^{\mathrm{i} \Phi_{\gamma}(t)} 
  = \frac{1}{N_{\gamma}}\sum_{i=1}^{N_{\gamma}} \e^{\mathrm{i} \phi_i^{\gamma}(t)},
  \label{equ:orderparameter_case4}
\end{eqnarray}
which represents the macroscopic collective oscillation of network $\gamma$.
Further,
the modulus $R_{\gamma}(t)$ indicates the degree of synchronization of network $\gamma$
and satisfies $0 \leq R_{\gamma} \leq 1$.
Using the order parameter, we can rewrite Eq.~(\ref{equ:dynamics_case4}) as
\begin{eqnarray}
  \dot{\phi}_i^{\gamma}(t)
  &=& \omega_i^{\gamma} + R_{\gamma} \sin(\Phi_{\gamma} - \phi_i^{\gamma} - \alpha)
  \nonumber \\
  & & + \epsilon R_{\nu} \sin(\Phi_{\nu} - \phi_i^{\gamma} - \beta).
  \label{equ:dynamics2_case4}
\end{eqnarray}

\par
We performed numerical simulations of Eq.~(\ref{equ:dynamics2_case4}) with Eq.~(\ref{equ:orderparameter_case4}).
In this study, the size of the network is $N_{A} = N_{B} = 2^{15}$.
The frequency dispersion is set to $w = \cos(\alpha) / 4$,
and the central values are set to $\omega_0^A = 1.0 \pi$ and $\omega_0^B = 1.01 \pi$.
For convenience,
we assumed $\omega_i^B = \omega_i^A + (\omega_0^B - \omega_0^A)$,
where $\omega_i^A$ is drawn from $g_{A}(\omega_i^A)$.
We set the coupling phase shifts $\alpha = \beta = 3\pi/8$ and the intensity of external coupling $\epsilon = 0.01$.
Under these conditions, each network is in a partially locked state,
where some of the oscillators in the network are synchronous and the others are asynchronous.
To illustrate the partially locked state,
Fig.~\ref{fig:Result_case4}(b) shows the snapshot of the phase oscillators in networks $A$ and $B$.
The oscillators are sorted in increasing order of their natural frequencies.
The aligned points in the center segment are synchronous oscillators that follow the collective phase.
The other scattered points are asynchronous oscillators that drift from the collective phase.

\par
In this case,
the theoretical method~\cite{nakao_phase_2018} cannot be applied
for reducing Eq.~(\ref{equ:dynamics_case4}) to Eq.~(\ref{equ:phase_description}),
because this model does not reflect the fully locked state that is the scope of application of the method.
Nevertheless,
we can obtain the dynamics of the collective phase
in the form of Eq.~(\ref{equ:phase_description}) with $\xi_{\gamma}(t) = 0$ analytically
using the Ott-Antonsen ansatz~\cite{ott_low_2008,ott_long_2009}.
The details of the derivation are given in Ref.~\cite{kawamura_phase_2010}.

\par
We apply the Bayesian inference method to extract the macroscopic phase coupling function.
While we observe the time-series of $\Phi_B$ from network $B$,
we observe the time-series of $\phi_i^A$ for $i = 10000, 20000, 30000$ or $\Phi_A$ from network $A$.
By changing the observed variable of network $A$,
we control the condition for the observed variable:
synchronous, asynchronous, or macroscopic collective oscillation.
The length of the observed time-series is up to $t = 1.0 \times 10^4$
and the sampling interval is set to $\Delta t = 0.1$.
The hyperparameters in Eq.~(\ref{equ:prior}) are set to $\alpha_0 = \beta_0 = 2.0$ and $\lambda_0 = 10$.
As in the cases of Sec.~\ref{subsec:case1} to Sec.~\ref{subsec:case3},
changing the observed variables causes a phase shift in the macroscopic phase coupling function.
Therefore,
to subtract the phase shift from each estimated model,
we calculate the argument of the complex number
$\langle \e^{\mathrm{i} (\Phi_A(t) - \phi_i^A(t))} \rangle_t$
as the phase shift between the observed variable $\phi_i^A$ and order parameter $\Phi_A$,
and then add it to the observed time-series of $\phi_i^A$.

\par 
Here, we explain the results of model identification for the observed variables $\phi_i^A~(i = 10000, 20000, 30000)$
using Figs.~\ref{fig:Result_case4}(c) and \ref{fig:Result_case4}(d).
Figure~\ref{fig:Result_case4}(c) shows
the time-series and distribution of $\phi_i^{\gamma} - \Phi_{\gamma}~(i = 10000, 20000, 30000)$
to explain the characteristics of each observed variable.
Figure~\ref{fig:Result_case4}(d) shows representative results of our method for each observed variable.
This figure also shows 
the theoretical curve (red dashed line) 
and a representative result for the observed variable $\Phi_A$ (blue dashed line).
Comparing the two lines,
we confirmed that the macroscopic phase coupling function could be extracted successfully from the time-series of the collective phases $\Phi_A$ and $\Phi_B$. 

\par
%% i = 10000
First, we explain the case of $i = 10000$.
Because $\phi_{10000}^A$ is a synchronous phase,
the time-series of $\phi_i^A - \Phi_A$ is almost constant
and the distribution $P(\phi_i^A - \Phi_A)$ is very narrow,
as shown in Fig.~\ref{fig:Result_case4}(c).
The yellow line in Fig.~\ref{fig:Result_case4}(d) is a representative result for $i = 10000$,
which is similar to the theoretical curve.
The result shown in Fig.~\ref{fig:Result_case4}(d) indicates that 
both directions of the macroscopic phase coupling function can be extracted
from the time-series of a synchronous oscillator and the collective phase.

\par
%% i = 20000
Second, $\phi_{20000}^A$ is the phase of an asynchronous oscillator.
As shown in Fig.~\ref{fig:Result_case4}(c),
the time-series of $\phi_i^A - \Phi_A$ increases monotonically,
and therefore, the time-series is distributed from $0$ to $2\pi$.
A representative result for $i = 20000$, i.e., the dark green line in Fig.~\ref{fig:Result_case4}(d),
shows the macroscopic phase coupling function $\hat{\Gamma}_{BA}$ (from $A$ to $B$) can be extracted,
whereas the opposite direction $\hat{\Gamma}_{AB}$ (from $B$ to $A$) cannot be extracted successfully.
This result indicates that the time-series of the collective phase of network $B$ is required
to successfully extract the macroscopic phase coupling function $\hat{\Gamma}_{BA}$,
whereas a synchronous phase of network $A$ is not required.

\par
%% i = 30000
Finally, $\phi_{30000}^A$ is also an asynchronous phase,
and it has a higher frequency than $\phi_{20000}^A$.
Figure~\ref{fig:Result_case4}(c) shows
the fast pace of the phase difference $\phi_i^A - \Phi_A$ and
the large variance of the distribution $P(\phi_i^A - \Phi_A)$.
A representative result for $i = 30000$, i.e., the light green line in Fig.~\ref{fig:Result_case4}(d),
shows that the shape of $\hat{\Gamma}_{BA}$ is slightly extracted, 
while $\hat{\Gamma}_{AB}$ is not at all.

\par
Here we explain our conclusion by returning to Fig.~\ref{fig:Result_case4}(a).
Let us consider the case where
the observed variable of network $B$ is the collective phase
and that of network $A$ is an asynchronous phase.
Under this condition,
the macroscopic phase coupling function from network $A$ to network $B$ (red arrow) can be extracted qualitatively
even when an asynchronous phase of network $A$ is observed.
In contrast,
the opposite direction of the macroscopic phase coupling function (gray arrow) cannot be extracted.
Therefore, 
we concluded that the difference in the type of observed variables of the two networks causes directional asymmetry in the robustness of extracting the macroscopic phase coupling function.
Although we did not investigate the accuracy of the model identification,
the results of this study show that
the pace of drift of the observed variable from the collective phase affects the accuracy.

%% conclusion %%%%%%%%%%%%%%%%%%%%%%%%%%%%%%%
\section{Discussion}
\label{sec:discussion}

\par
We extended the range of applications of the Bayesian inference method~\cite{ota_direct_2014}
from a phase representing the state of an oscillator
to a macroscopic phase representing the state of an entire network with collective oscillation~\cite{
  nakao_phase_2018, ott_low_2008, ott_long_2009, kawamura_phase_2010},
and demonstrated the Bayesian method
in four cases (Secs.~\ref{subsec:case1}, \ref{subsec:case2}, \ref{subsec:case3}, and \ref{subsec:case4})
to extract macroscopic phase coupling functions, 
which describe the synchronization mechanism between networks,
directly from time-series data.
In Sec.~\ref{subsec:case1} to Sec.~\ref{subsec:case3},
we focused on the networks with collective oscillation in a fully locked state.
In Sec.~\ref{subsec:case1},
we considered the case of two networks of FHN elements,
which have both excitable and oscillatory elements.
In Sec.~\ref{subsec:case2},
we considered the case of three networks of FHN elements
as a generalization of Sec.~\ref{subsec:case1}
in terms of the number of networks and structures of internal couplings.
In Sec.~\ref{subsec:case3},
we investigated the case of two networks of van der Pol oscillators.
In these cases,
we considered the three types of observed variables: one excitable element, one oscillatory element, and a mean-field,
to evaluate how the choice of the observed variables affects the statistical inference for the phase coupling functions.
Further,
we investigated the case of two networks of globally coupled phase oscillators in a partially locked state in Sec.~\ref{subsec:case4}.
We revealed the possibility of extraction of the macroscopic phase coupling function
from the time-series of an asynchronous oscillator.

\par
In the cases of the fully locked state,
our results showed that the same phase coupling function can be extracted from the time-series of any one element in each network, 
as well as from the time-series of each network's mean-field.
These results indicate that
the statistical inference for the macroscopic phase coupling function is not affected by the choice of the observed variables;
thus, we can consistently extract the macroscopic phase coupling function between networks.
In addition,
we extracted the macroscopic phase coupling function between networks consistently
regardless of the two transformation methods from the observed time-series to the phase time-series, i.e.,
linear interpolation for networks of FHN elements (Secs.~\ref{subsec:case1} and \ref{subsec:case2})
and the Hilbert transform for networks of van der Pol oscillators (Sec.~\ref{subsec:case3}).
We should remark that
the assumption of the fully locked state enables us
to extract the macroscopic phase coupling function from the time-series of an excitable element,
although the conventional phase reduction theory cannot apply to an excitable element.
The ability to extract the macroscopic phase coupling function
regardless of the transformation method to a phase time-series or the types of observed variables
is useful in experimental situations.

\par
In another case, the partially locked state,
our results showed that the macroscopic phase coupling function can be extracted
even when an asynchronous oscillator is observed.
This result reveals that
the difference in the type of observed variables of the two networks causes directional asymmetry in the robustness of extracting the macroscopic phase coupling function.
Let us consider the situation illustrated in Fig.~\ref{fig:Result_case4}(a), 
where an asynchronous oscillator in network $A$ and the macroscopic collective oscillation of network $B$ are observed.
We were able to extract the macroscopic phase coupling function from network $A$ to network $B$ more robustly than in the opposite direction.
Therefore, 
to extract the macroscopic phase coupling function from network $A$ to network $B$,
the time-series of the macroscopic collective oscillation of network $B$ is essential,
whereas the observed variable of network $A$ can be an asynchronous oscillator.
However, the extraction fails if the observed asynchronous oscillator of network $A$ has a fast pace of drift from collective oscillation.
Therefore, the validation of extracting the macroscopic phase coupling function
from the time-series of synchronous and asynchronous oscillators remains as a further study.

\par
Combining the Bayesian inference and the concept of collective oscillation
may provide a significant insight into the analysis of the macroscopic phase coupling function between networks.
As mentioned in Sec.~\ref{sec:intro},
there are various ways to observe the dynamics of brain regions for analyzing coupling functions between them.
However, the real meaning of the coupling function extracted from the time-series of the spike signals of one neuron is unclear. 
Our results suggest the possibility that the time-series of an asynchronous neuron spike in the sender region of the coupling can be used, instead of LFP, 
to extract the macroscopic phase coupling function between neural populations with collective oscillations.
However, the time-series of LFP in the receiver region of the coupling is necessary.
Arguably, this suggestion can apply to various domains, 
because the only assumption for our method is one of two types of collective oscillation: fully locked state or partially locked state.

\par
In this study,
we considered a macroscopic phase description where each network has collective oscillation,
as described in Refs.~\cite{nakao_phase_2018, ott_low_2008, ott_long_2009, kawamura_phase_2010}.
In particular, the phase description for the fully locked state~\cite{nakao_phase_2018}
has been generalized to rhythmic spatiotemporal dynamics, 
such as reaction-diffusion systems~\cite{nakao_phase-reduction_2014}.
Further,
some other theoretical frameworks for a network phase response analysis have been developed:
phase coherent states in globally coupled noisy identical oscillators~\cite{
  kawamura_collective_2008, kawamura_collective_2014, kawamura_phase_2010-1} 
and fully phase-locked states in networks of coupled nonidentical elements~\cite{
  kori_collective-phase_2009, kawamura_phase_2014}.
Particularly,
the framework developed in Ref.~\cite{kawamura_collective_2017}
is a generalization for a network of globally coupled noisy identical oscillators
to address excitable elements and strong internal couplings.
This type of generalization corresponds to the assumption we applied for the fully locked state in this study.
We will attempt to extend the range of application of the Bayesian inference method~\cite{ota_direct_2014}
for the aforementioned situations in future research.

\par 
We used one of the Bayesian inference methods~\cite{ota_direct_2014}
to extract the macroscopic phase coupling functions in Kuramoto--Daido form.
However, other statistical inference methods~\cite{
  kiss_predicting_2005, tokuda_inferring_2007, miyazaki_determination_2006,
  galan_efficient_2005, ota_weighted_2009, imai_improvement_2016, ermentrout_relating_2007,
  kralemann_uncovering_2007, kralemann_phase_2008,
  kralemann_vivo_2013, kralemann_reconstructing_2011, kralemann_reconstructing_2014,
  stankovski_inference_2012, duggento_dynamical_2012, stankovski_coupling_2017, jafarian_structure_2019, rosch_calcium_2018}
should also be able to extract the macroscopic phase response.
Finally, we mention some advantages of adopting the Kuramoto--Daido form for statistical extraction of phase coupling functions.
To extract the terms of high-order harmonics in the pairwise phase coupling function $\hat{\Gamma}_{ij}$ up to the $M$th-order,
the Kuramoto--Daido form requires the $\mathcal{O}(M)$ model parameter.
This form requires lower computation costs than the general form used in Refs.~\cite{
  kralemann_uncovering_2007, kralemann_phase_2008,
  stankovski_inference_2012, duggento_dynamical_2012}.
Another advantage is that 
the extraction of the phase coupling function in the Kuramoto--Daido form can be applied
to the phase time-series reconstructed from the event sampling measurement.
Linear interpolation is a representative method to reconstruct the phase time-series from event sampling measurements~\cite{
  ota_interaction_2020, suzuki_bayesian_2018}
and this method reflects the perturbation in averaged form over one period.
The Kuramoto--Daido form is suitable for this situation
because it reflects the average of the perturbations that arise from the coupling function over one period of oscillation.

\par
In addition to these advantages, the Kuramoto--Daido form can be extended to various cases.
As mentioned in Sec.~\ref{sec:method},
the $n:m$ phase coupling function can be adopted to evaluate $n:m$ synchronization~\cite{onojima_dynamical_2018}.
Moreover,
there are some forms to describe the higher-order approximation with respect to perturbation~\cite{battiston_networks_2020}.
Indeed, to extract the accurate phase dynamics,
it is necessary to consider triplet or $N$-body phase coupling functions~\cite{
  kralemann_reconstructing_2011, kralemann_reconstructing_2014,
  duggento_dynamical_2012, stankovski_inference_2012, stankovski_coupling_2015} 
for the higher-order terms,
even when only direct pairwise couplings exist.
However, the phase equation requires only pairwise phase coupling functions
since the first-order approximation is sufficient for weak direct pairwise coupling.

%% acknowledgements %%%%%%%%%%%%%%%%%%%%%%%%%%%%%%%
\begin{acknowledgments}
This work was supported by JSPS (Japan) KAKENHI Grant Numbers JP20K21810, JP20K20520, JP20H04144, JP20K03797, JP18H03205, JP17H03279.
\end{acknowledgments}

%% acknowledgements %%%%%%%%%%%%%%%%%%%%%%%%%%%%%%%
\appendix
\section{Supplemental information for Case~2 and Case~3}
\label{sec:appendix}

\par
\begin{figure*}
  \includegraphics[width=0.9\textwidth]{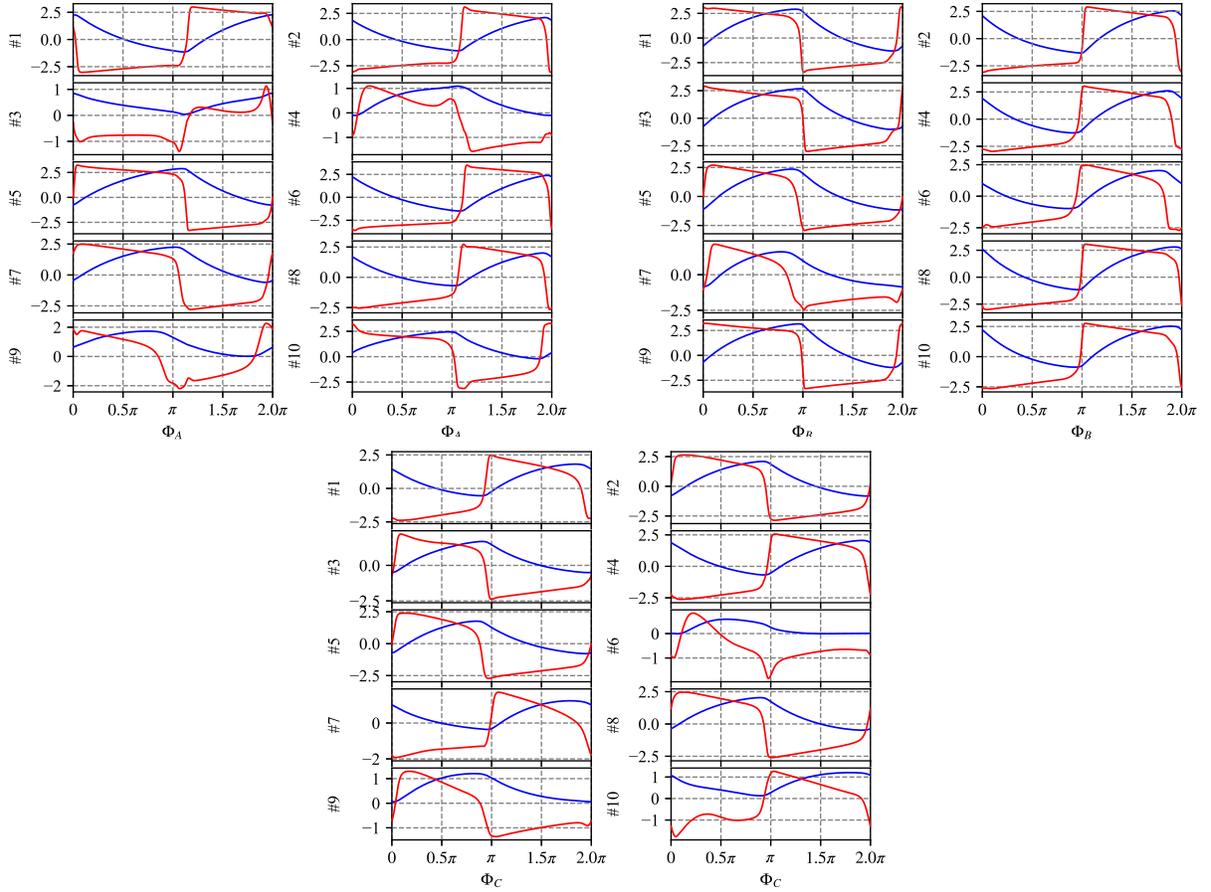}
  \caption{
    Waveforms of networks $A$ (top left), $B$ (top right), and $C$ (lower middle)
    with collective oscillation in Case~2.
    Each panel shows $u_i^{\gamma}(\Phi_{\gamma})$ (blue) and $v_i^{\gamma}(\Phi_{\gamma})$ (red)
    for one period of element $i~(i = 1, 2, \ldots, 10)$ in network $\gamma \in \{A, B, C\}$.
  }
  \label{fig:Append_case2}
\end{figure*}

Figure~\ref{fig:Append_case2} shows the waveforms for one period
of each network with collective oscillation in Case~2.
The frequencies of the collective oscillations of the three networks are
$\Omega_A \simeq 0.159$, $\Omega_B \simeq 0.148$, and $\Omega_C \simeq 0.172$.

\par
\begin{figure*}
  \includegraphics[width=0.9\textwidth]{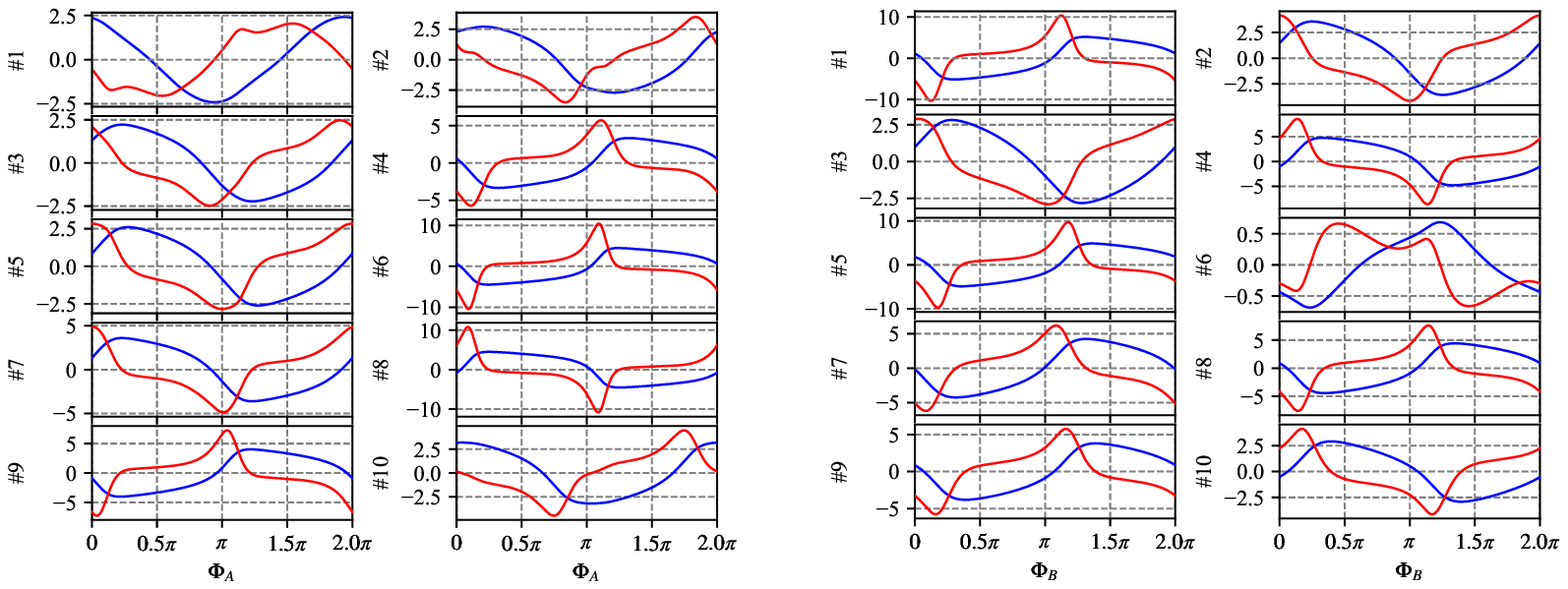}
  \caption{
    Waveforms of networks $A$ (left) and $B$ (right)
    with collective oscillation in Case~3.
    Each panel shows $u_i^{\gamma}(\Phi_{\gamma})$ (blue) and $v_i^{\gamma}(\Phi_{\gamma})$ (red)
    for one period of element $i~(i = 1, 2, \ldots, 10)$ in network $\gamma \in \{A, B\}$.
  }
  \label{fig:Append_case3}
\end{figure*}

Figure~\ref{fig:Append_case3} shows the waveforms for one period
of each network with collective oscillation in Case~3.
The frequencies of the collective oscillations of the two networks are
$\Omega_A \simeq 0.900$ and $\Omega_B \simeq 0.939$.
%%

%% reference %%%%%%%%%%%%%%%%%%%%%%%%%%%%%%%
% \bibliography{001_Collective_Oscillation,001_add}


\end{document}